\documentclass[a4paper,11pt]{article} 
\pdfoutput=1
\usepackage{jheppub} 
\usepackage[utf8x]{inputenc} 
\usepackage[T1]{fontenc} 
\usepackage{bbm,bm,commath,mathtools,wasysym,xfrac} 
\usepackage{xcolor} 
\usepackage{trimclip} 
\usepackage[nottoc,notlof,notlot]{tocbibind} 

\usepackage{comment} 
\usepackage{tikz} 
\usepackage{tikz-cd} 
\usepackage{tcolorbox}
\usepackage{float}
\usepackage{slashed}
\graphicspath{{figures/}} 

\DeclareUnicodeCharacter{"0393}{\Gamma}
\DeclareUnicodeCharacter{"0394}{\Delta}
\DeclareUnicodeCharacter{"0398}{\Theta}
\DeclareUnicodeCharacter{"039B}{\Lambda}
\DeclareUnicodeCharacter{"039E}{\Xi}
\DeclareUnicodeCharacter{"03A0}{\Pi}
\DeclareUnicodeCharacter{"03A3}{\Sigma}
\DeclareUnicodeCharacter{"03A5}{\Upsilon}
\DeclareUnicodeCharacter{"03A6}{\Phi}
\DeclareUnicodeCharacter{"03A8}{\Psi}
\DeclareUnicodeCharacter{"03A9}{\Omega}
\DeclareUnicodeCharacter{"03B1}{\alpha}
\DeclareUnicodeCharacter{"03B2}{\beta}
\DeclareUnicodeCharacter{"03B3}{\gamma}
\DeclareUnicodeCharacter{"03B4}{\delta}
\DeclareUnicodeCharacter{"03B5}{\epsilon}
\DeclareUnicodeCharacter{"03B6}{\zeta}
\DeclareUnicodeCharacter{"03B7}{\eta}
\DeclareUnicodeCharacter{"03B8}{\theta}
\DeclareUnicodeCharacter{"03D1}{\vartheta}
\DeclareUnicodeCharacter{"03B9}{\iota}
\DeclareUnicodeCharacter{"03BA}{\kappa}
\DeclareUnicodeCharacter{"03BB}{\lambda}
\DeclareUnicodeCharacter{"03BC}{\mu}
\DeclareUnicodeCharacter{"03BD}{\nu}
\DeclareUnicodeCharacter{"03BE}{\xi}
\DeclareUnicodeCharacter{"03C0}{\pi}
\DeclareUnicodeCharacter{"03C1}{\rho}
\DeclareUnicodeCharacter{"03C3}{\sigma} 
\DeclareUnicodeCharacter{"03C4}{\tau}
\DeclareUnicodeCharacter{"03C5}{\upsilon}
\DeclareUnicodeCharacter{"03C6}{\phi}
\DeclareUnicodeCharacter{"03D5}{\varphi}
\DeclareUnicodeCharacter{"03C7}{\chi}
\DeclareUnicodeCharacter{"03C8}{\psi}
\DeclareUnicodeCharacter{"03C9}{\omega}
\DeclareUnicodeCharacter{"21D0}{\Leftarrow}
\DeclareUnicodeCharacter{"0212B}{\AA}
\DeclareUnicodeCharacter{"00B7}{\cdot}
\DeclareUnicodeCharacter{"00B0}{^{\circ}}
\DeclareUnicodeCharacter{"266A}{\eighthnote}
\DeclareUnicodeCharacter{"266B}{\twonotes}
\PrerenderUnicode{¹}\PrerenderUnicode{²}
\PrerenderUnicode{ⁿ}\PrerenderUnicode{×}

\newcommand\eqst[1] {\begin{multline}#1\end{multline}}

\definecolor{darkgreen}{rgb}{0,0.5,0}

\newcommand\cmnt[1] {{\color{red}#1}}

\renewcommand{\beforetochook}{\bigskip\bigskip} 
\numberwithin{equation}{section} 
\begin{document}
\title{\texorpdfstring{Recursion relations for scattering amplitudes with massive particles II: massive vector bosons}}

\author[a]{Sourav Ballav}
\affiliation[a]{Indian Institute of Science Education and Research Bhopal, \\
Bhopal Bypass Rd, Bhauri, Madhya Pradesh 462066, India\\
}
\emailAdd{souravballav@gmail.com} 
\author[b]{and Arkajyoti Manna}
\emailAdd{arkajyotim@imsc.res.in}
\affiliation[b]{The Institute of Mathematical Sciences \\
Homi Bhabha National Institute (HBNI)\\
IV Cross Road, C.~I.~T.~Campus, \\
  Taramani, Chennai, 600113  Tamil Nadu, India \\}

\abstract{Using the recently introduced recursion relations with covariant massive-massless shift,  we study tree-level scattering amplitudes involving a pair of massive vector bosons and an arbitrary number of gluons in the massive spinor-helicity formalism.  In particular, we derive  compact expressions for  cases in which i) all gluons are of the same helicity and ii) one gluon has flipped helicity and is colour adjacent to one of the massive particles.
We provide numerous consistency checks of our results including the exact match of high energy limits with well known MHV and NMHV amplitudes in pure Yang-Mills theory.  As a corollary, we obtain an alternative novel representation of the NMHV amplitude.}


\maketitle
\section{Introduction}
Spinor-helicity formalism has revolutionized our understanding of the S-matrix theory for massless particles. By trivialising the non-linear constraints such as the Gram determinant condition that Mandelstam variables have to satisfy,  expressing the external scattering data in terms of spinor-helicity variables leads to remarkably simpler and conceptually revealing expressions for scattering amplitudes.  For example, the Parke-Taylor amplitude \cite{PhysRevLett.56.2459} took a strikingly simple form when the external momenta and polarisation data were expressed in terms of spinors \cite{GUNION1985333,  Xu:1986xb, Gastmans:1990xh,  Dixon:1996wi}.  As spinors are complex, the real power of this formalism was revealed when complex deformations of external momenta was used to derive on-shell recursion relations by \citep{Britto:2004ap,Britto:2005fq}. In fact,  computations of the tree-level amplitudes in gauge theories and gravity get immensely simplified by implementing the BCFW recursion relations in the spinor-helicity formalism.

The recursion relations construct higher-point amplitudes in terms of lower-point amplitudes while staying on-shell.   These recursion relations were generalized to the case of massive particles in \citep{Schwinn:2005pi, Franken:2019wqr, Badger:2005zh, Badger:2005jv, Ferrario:2006np}. In \citep{Schwinn:2005pi, Franken:2019wqr} multiple massive momenta were complexified to study the scattering amplitudes.  However, in these works,  the massive momenta were written in terms of certain light-like momenta due to which the covariance (with respect to the little group action of external particles) was broken.  In \citep{ Badger:2005zh} tree-level amplitudes with a pair of  massive scalars and up to four gluons were computed using the BCFW shift on a pair of massless external particles (gluons) and later this method was used to compute several lower-point tree-level amplitudes involving fermions and massive vector bosons (spin-1) scattering with gluons \citep{Badger:2005jv}.  In an another development \citep{Ferrario:2006np}, tree-level amplitude of a pair of complex scalar and an arbitrary number of positive helicity gluons was computed using the Berends-Giele and on-shell recursion relations, and obtained an extremely compact expression. In \citep{Schwinn:2007ee} this was further extended to compute the amplitude involving a pair of massive quarks and arbitrary number of gluons.

Recently, in a remarkable work a little group covariant spinor-helicity formalism for massive particles was introduced \citep{Arkani-Hamed:2017jhn}.  In a beautiful paper, Ochirov combined this 
formalism with the recursion relations proposed in \citep{Ochirov:2018uyq}.  In particular using the BCFW shift on a pair of gluons, Ochirov derived formulae for two classes of $n$-point amplitudes involving massive quarks,  consistent with the previous results in \citep{Schwinn:2007ee, Ferrario:2006np}.

In \citep{Ballav:2020ese, Aoude:2019tzn,Falkowski:2020aso},  a new set of recursion relations were derived in the massive spinor-helicity formalism for on-shell amplitudes by complexifying one massive and one massless external states. 
These complex momentum shifts (involving a complex parameter $z$) were then realized in the spinor-helicity basis by considering little group covariant deformations of massive and massless spinor-helicity variables.
We call this particular shift as the covariant massive-massless shift and refer to the resulting recursion as the covariant recursion relations.  In earlier work \citep{Ballav:2020ese},  we  used these recursion relations to study tree-level lower-point amplitudes in scalar QCD as well as amplitudes involving massive vector bosons in the Higgsed Yang-Mills theory.  We also classified all of valid covariant massive-massless shifts for these theories by requiring that the amplitude does indeed vanish as the complex deformation parameter $z$ tends to $\infty$.  
In this paper, we further use these recursion relations to compute tree-level $n$-point amplitudes in the Higgsed Yang-Mills theory (that includes massive spin-$1$ particles and gluons as quanta of the theory).


 As is well known, one of the earliest and striking applications of the BCFW recursion technique was in (1) the proof of Parke-Taylor formula for $n$-point maximally helicity violating(MHV) amplitudes and (2) the ease with which tree-level next-to-maximally helicity violating (NMHV)  amplitudes could be computed. The power of BCFW recursion technique could be seen from the fact that, the n-particle MHV and NMHV amplitudes can be obtained by using a single recursion.  Completely analogously,  we compute two classes of $n$-point amplitudes involving a pair of massive vector bosons and gluons in the external data such that in the high energy limit, these amplitudes reduce to MHV and NMHV gluon amplitudes respectively.

 The paper is organised as follows. In section \ref{section-2},  we review the massive spinor-helicity formalism and the covariant recursion relations.  In section \ref{section-3}, we compute the tree-level  colour-ordered amplitude involving a pair of massive vector bosons and an arbitrary number of gluons of same helicity that is massive analogue of the MHV amplitude. To obtain this amplitude we  first use a simple relation between the amplitude involving two massive vector bosons and $(n-2)$ positive helicity gluons, and the amplitude involving two massive scalars and $(n-2)$ positive helicity gluons. This relation is a covariantized version of a relation that has appeared in \cite{Boels:2011zz} for a particular choice of spin projection of the massive particles.  We then prove this result by using the method of induction and the covariant recursion relation. We also check consistency of this amplitude by taking the high energy limit which exactly matches with the pure gluon MHV amplitude.

 In section \ref{section-4}, we turn to the main focus of this paper which is the computation of the tree-level colour-ordered amplitude involving a pair of massive vector bosons,  one negative helicity gluon that is colour-adjacent to one of these massive particles and an arbitrary number positive helicity gluons. We obtain this amplitude using the covariant recursion relations as proposed in \citep{Ballav:2020ese}.  We will find that the single covariant recursion involves only subamplitudes that have been previously computed.
 Finally we check the consistency of this result by taking the high energy limit and produce the $n$-point NMHV amplitude. 
 
 We conclude in section \ref{summary} with a short summary and outline some immediate future directions, and collect some technical materials in the appendices.

\section{Review of covariant recursion relation}\label{section-2}
Scattering amplitudes are Lorentz invariant objects and transform covariantly under little group which is ISO(2) for massless particles and SU(2)  for massive particles in four dimension.  Hence  we label massless states by the helicity ($h$) of the particle and use symmetric $2S$ representation of SU(2) to represent the massive spin-$S$ one-particle state,  instead of using the standard representation of SU(2) introducing a preferred $z$-direction which breaks the rotational invariance of S-matrix.  
Amplitude involving a massless particle with momentum $p_j$ and a massive particle with momentum $p_i$ and spin $S$ then transforms under little group as follows \citep{Arkani-Hamed:2017jhn}
\begin{align}
\mathcal{A}^{h}_{I_1I_2...I_{2S}}\left(t \lambda _j,t^{-1}\tilde{\lambda}_j;W\lambda_i, W^{-1}\tilde{\lambda}_i; \cdots \right)\rightarrow t^{-2h} &W_{I_1}{}^{J_1} W_{I_2}{}^{J_2}\cdots W_{I_{2S}}{}^{J_{2S}}\cr 
& \mathcal{A}^{h}_{J_1J_2...J_{2S}} \left(\lambda _j,\tilde{\lambda}_j;\lambda_i, \tilde{\lambda}_i;\cdots\right)\,,
\end{align} 
where $t$ is a complex number associated to the $j$-th massless particle and $W$'s are SU(2) matrices in the fundamental representation  associated to the $i$-th massive particle.  Since the amplitude is covariant in little group indices,  it is useful to express this directly in terms of functions which transform covariantly under the little group transformations.  In four dimensions, the well-known choice is to use the  ``spinor-helicity variables''.  As we will review in the subsequent section,  the spinor-helicity variables ($\lambda ,\tilde{\lambda}$) are functions of on-shell momentum of the particle and transform covariantly under the little group transformations.

\subsection{Spinor-helicity formalism in four dimensions}
The basic goal of this formalism is to express on-shell momentum in terms of spinor-helicity variables.  To introduce these variables,  we consider the SL(2,$\mathbb{C}$) representation of momentum $4$-vector( $p_\mu \sigma ^\mu_{\alpha \dot{\alpha}}=p_{\alpha \dot{\alpha}}$).  For massless particles,  this is a rank-$1$ matrix and can be expressed as
\begin{align}
p_{\alpha \dot{\alpha}}=\lambda _\alpha \tilde{\lambda}_{\dot{\alpha}}\,,
\end{align}
where $\lambda _\alpha$ and $\tilde{\lambda}_{\dot{\alpha}}$ are two-component Weyl spinors, known as massless spinor-helicity variables.  Since we can always rescale the spinor-helicity variables
\begin{align}
\lambda _\alpha \longrightarrow t\lambda _\alpha\,, \qquad \tilde{\lambda}_{\dot{\alpha}} \longrightarrow t^{-1}\tilde{\lambda}_{\dot{\alpha}} \,,
\end{align}
it is impossible to assign unique spinor-helicity variables to express $p_{\alpha \dot{\alpha}}$.  But this scaling is exactly the little group scaling for massless particle.  Thus we identify $\lambda _\alpha$ and $\tilde{\lambda}_{\dot{\alpha}}$ as objects having little group weight $\pm 1$ respectively. Using spinor-helicity variables, we define Lorentz invariant and little group covariant angle and square brackets as
\begin{align}
\langle ij\rangle :=\lambda _i^\alpha \lambda _{j\alpha}\,, 
\qquad [ij]:=\tilde{\lambda}_{i\dot{\alpha}}\tilde{\lambda}_j^{\dot{\alpha}} \,, \qquad 2 p\cdot q=\langle pq\rangle [qp] \,.
\end{align}
These brackets are the basic building blocks of scattering amplitude in spinor-helicity formalism.  Massless spinor-helicity variables satisfy the Weyl equation 
\begin{align}
p_i|i\rangle =p_i|i]=0\,.
\end{align}

Next we turn to the particles with mass.  In this case det$(p_{\alpha \dot{\alpha}})=p^\mu p_\mu \neq 0$.  Hence $p_{\alpha \dot{\alpha}}$ is expressed as a linear combination of two rank-$1$ objects \citep{Arkani-Hamed:2017jhn}
\begin{align}
p_{\alpha \dot{\alpha}}=\sum _{I,J=1}^2 \epsilon _{IJ} \lambda ^I_\alpha \tilde{\lambda}^J_{\dot{\alpha}} \,,
\end{align}
where $(I,J)$ are SU(2) little group indices for massive particle.  The variables $\lambda ^I_\alpha , \tilde{\lambda}^J_{\dot{\alpha}}$ are called massive spinor-helicity variables .  Similar to the massless case,  there is no unique way to fix these spinors, satisfying the above relation due to the following transformation
\begin{align}
\lambda ^I _\alpha \longrightarrow W^I~_J \lambda ^J_\alpha \, \qquad \tilde{ \lambda}^J_{\dot{\alpha}} \longrightarrow (W^{-1})^J~_K \tilde{\lambda}^K_{\dot{\alpha}} \,.
\end{align}
Unlike the massless case,  these transformations do not correspond to the little group transformation of massive particle as $W$ can be any GL(2) 
matrix.  But if we demand det$(\lambda ^I_\alpha)=$ det $(\tilde{\lambda}^J_{\dot{\alpha}})=m$, then it can be shown that $W$ is indeed a SU(2) matrix for real momenta, reflecting the above transformations as little group transformation.  

The massless spinor-helicity variables $\lambda _\alpha ,  \tilde{\lambda}_{\dot{\alpha}}$, which satisfy Weyl equation are  independent of each other.  However, the dotted and undotted massive spinor-helicity variables are related to each other via Dirac equation
\begin{align}
p_{\alpha \dot{\alpha}}\lambda ^\alpha _I=-m\tilde{\lambda}_{I\dot{\alpha}}~;\quad p_{\alpha \dot{\alpha}} \tilde{\lambda}^{\dot{\alpha}}_I=m\lambda _{I\alpha}\,.  \label{diraceqn}
\end{align}
Therefore the scattering amplitude involving massive particles can be expressed in terms of only either $\lambda ^\alpha _I$ or $\tilde{\lambda}_{I\dot{\alpha}}$ as opposed to amplitude with only massless particles. This feature of amplitude proves extremely useful to classify all possible three-particle amplitudes \citep{Arkani-Hamed:2017jhn} involving massive as well as massless particles.

\subsection{Three-particle amplitude}
In this section, we briefly review all the required three-particle amplitudes which will be used as basic building blocks to construct higher-point amplitudes using recursion scheme.  Due to kinematics,  the three-particle amplitude involving massless particles with helicity $h_{1,2,3}$ can be expressed either in terms of angle or square brackets.  Little group transformation then fixes the structure upto an overall multiplicative constant, 
\begin{align}
\mathcal{A}_3^{h_1h_2h_3}[1,2,3]&=g [12]^{h_1+h_2-h_3}[23]^{h_2+h_3-h_1}[31]^{h_3+h_1-h_2}~ ;\quad \text{with $h_1+h_2+h_3 >0$}\nonumber \\
&= g^{'}\langle 12\rangle ^{h_3-h_1-h_2} \langle 23\rangle ^{ h_1-h_2-h_3} \langle 31\rangle ^{h_1-h_2-h_3}~ ;\quad \text{with $h_1+h_2+h_3 <0$}\,,
\end{align} 
with two distinct representations ensuring smooth vanishing limit in Minkowski signature as individual spinor products vanish in this signature for real momenta. 

Three-particle amplitudes involving two massive particles with mass $m$ and spin-$1$ coupled with a massless particle of helicity $h$
are given by \citep{Arkani-Hamed:2017jhn}
\begin{align}
\mathcal{A}_3^{+h}(\textbf{1},\textbf{2},3^h)=\frac{g}{m}x_{12}^h \langle \textbf{1}\textbf{2}\rangle ^{2}\,;\quad \mathcal{A}_3^{-h}(\textbf{1},\textbf{2},3^{-h})=\frac{g}{m}x_{12}^{-h} [ \textbf{1}\textbf{2}]^{2}\,.  \label{minimalthreepoint}
\end{align}
These amplitudes have well-behaved massless limit and relevant for the Higgsed Yang-Mills theory \cite{johansson19}.
Here $x_{12}$ is a non-local factor arises due to the degeneracy of masses.  It is defined as follows
\begin{align}
x_{12}=\frac{\langle \zeta |p_1|3]}{m\langle \zeta 3\rangle} \quad \text{or}\quad x_{12}^{-1}=\frac{\langle 3|p_1|\zeta]}{m[3\zeta]}~,
\end{align} 
where $\zeta $ is a reference spinor. We denoted massive spinor-helicity variables in bold notation and omitted little group indices in the amplitude.  The angle and square brackets of these bolded variables are defined as a symmetric product of spinor brackets in $SU(2)$ indices.  For example, 
\begin{align}
\langle \textbf{1}\textbf{2}\rangle ^2&= \langle 1^{I_1}2^{J_1}\rangle \langle 1^{I_2}2^{J_2}\rangle +\langle 1^{I_2}2^{J_1}\rangle \langle 1^{I_1}2^{J_2}\rangle \,,\\
\langle 3\textbf{2}\rangle ^2 &=\langle 32^{J_1}\rangle \langle 32^{J_2}\rangle \,.
\end{align} 
All the amplitudes that we are going to discuss in this note are obtained by gluing three-point amplitudes involving massive spin-1 particles.  So we can use these three-particle amplitudes as basic building blocks.

\subsection{Massive-massless shift and the covariant recursion }
\label{mmshift}
We use the two-line little group covariant massive-massless shift introduced in \cite{Ballav:2020ese,  Aoude:2019tzn} to compute four- and higher-particle amplitudes involving gluons and massive vector bosons in the Higgsed Yang-Mills theory\footnote{It is a gauge theory that describes the interaction between massive vector bosons and gluons. The three-point interaction between massive spin-1 particles and gluon in this theory can be obtained by Higgsing a theory of a scalar field coupled to SU(2) Yang-Mills field and an abelian gauge field \cite{johansson19}. } .  Although this particular shift of external momenta is in the similar spirit with the well known BCFW shift but involves complex deformation of massless and as well as massive momenta.  Let us consider that the massive and massless momenta,  denoted by $p_i$ and $p_j$ respectively,  are analytically continued to the complex plane while staying on-shell
\begin{align}
p_i \longrightarrow \widehat{p}_i^\mu =p_i^\mu -zq^\mu\,, \qquad p_j \longrightarrow \widehat{p}_j^\mu =p_j^\mu +zq^\mu\,.\label{momentumshift}
\end{align}
Here $z$ is complex deformation parameter and $q^\mu$ is lightlike shift vector satisfying the following conditions
\begin{align}
q\cdot p_i=0=q\cdot p_j\,.
\end{align} 
The momentum shift can be achieved by the following deformations of the massive and massless spinor-helicity variables \cite{Ballav:2020ese}
\begin{align}
\text{massive shift }:& \qquad	\widehat{\lambda}^I_{i\alpha}=\lambda ^I_{i\alpha}  ~,\qquad \widehat{\tilde{\lambda}}^I_{i\dot{\alpha}}=\tilde{\lambda} _{i\dot{\alpha}}^I-\frac{z}{m}\tilde{\lambda}_{j\dot{\alpha}}[i^Ij]\,,
	\label{massiveshift}\\
\text{massless shift }: & \qquad    \widehat{\tilde{\lambda}}_{j\dot{\alpha}} =\tilde{\lambda}_{j\dot{\alpha}}~, \qquad \widehat{\lambda}_{j\alpha}=\lambda _{j\alpha} +\frac{z}{m}p_{i\alpha \dot{\beta}}\tilde{\lambda}_j^{\dot{\beta}}\,.\label{masslessshift}
\end{align}
In \citep{Ballav:2020ese}, all the valid covariant massive-massless shifts in the Higgsed Yang-Mills theory have been classified. These are denoted by $[\textbf{m}+\rangle$ and $[-\textbf{m}\rangle$. The shifts in \eqref{massiveshift} and \eqref{masslessshift} are of the type$[\textbf{m}+\rangle$ and denoted as $[\textbf{i}j^+\rangle$.
 Here $\pm$ indicate the helicity of the massless particle and $m$ denotes the mass of the massive particle.
In particular, it has been shown that the deformed amplitude vanishes as the deformation parameter $z$ tends to $\infty$.  This proof is quite general in the sense that it does not depend number of external particles as long as one can deform a single massive and massless external momenta.  
We use this in the following recursion relation to compute amplitudes involving gluons having arbitrary helicity and massive vector bosons
\begin{align}
\mathcal{A}_n=\sum _{I} \widehat{\mathcal{A}}_{l+1}(z_I) \frac{1}{P^2-m^2}\widehat{\mathcal{A}}_{r+1} (z_I)+\sum _{J} \widehat{\mathcal{\tilde{A}}}_{l+1}(z_J) \frac{1}{P^2}\widehat{\mathcal{\tilde{A}}}_{r+1} (z_J)\,.\label{recursion}
\end{align}
The sum includes all possible scattering channels as well as spin or helicity states of the exchange particle.
We consider only colour-ordered amplitudes instead of fully colour-dressed tree-level amplitudes since the latter can be constructed from the former using the well known colour decomposition \citep{DelDuca:1999rs, Dixon:1996wi, Johansson:2015oia, Ochirov:2019mtf, Maltoni:2002mq, Melia:2015ika}.


\section{Scattering of massive vector bosons with positive helicity gluons }\label{section-3}
One of the earliest applications of the BCFW recursion relations was to provide an extremely simple proof of the formula for the $n$-point  MHV amplitude using the principle of induction.   As alluded to in the introduction, we wish to similarly apply the covariant recursion relations for the case of amplitudes involving massive particles.  
In this section, we consider an $n$-point amplitude involving a pair of colour adjacent massive vector bosons and $(n-2)$ positive helicity gluons.  This particular scattering amplitude serves as a massive analogue of the $n$-point MHV amplitude,  as we will see later in the following section that it reproduces the $n$-point MHV amplitude in the high energy limit.  We will show that this massive vector boson amplitude can also be derived inductively in the covariant recursion scheme, just as the $n$-point MHV amplitude was derived using BCFW.

 The $n$-point MHV amplitude was already derived in \cite{PhysRevLett.56.2459} before the discovery of BCFW recursion.  But using the BCFW recursion, the derivation became extremely simple.  In our case, the massive vector boson amplitude that we want to compute using the covariant recursion is not known.  To obtain this amplitude, we adopt a different strategy: firstly, we relate this amplitude to a known amplitude involving a pair of massive scalars and $(n-2)$ positive helicity gluons by using the little group covariant version of a formula that first appeared in \cite{Boels:2011zz}.  Secondly, we prove this formula in detail by making use of the covariant recursion relations and the principle of induction. 

The relation between the $n$-particle amplitude involving a pair of massive bosons and positive helicity gluons and the $n$-particle amplitude involving pair of massive scalars and positive helicity gluons is the following
\begin{align}
\mathcal{A}_{n}[\textbf{1},2^+,\ldots, (n-1)^+, \textbf{n}]=\frac{\langle \textbf{1}\textbf{n}\rangle ^2}{m^2}\mathcal{A}_{n}[\textbf{1}^0,2^+,\ldots,  (n-1)^+, \textbf{n}^0]\,.\label{vector-scalar}
\end{align}
This is a covariantization (in little group indices) of a relation  that has appeared previously in \citep{Boels:2011zz} for a particular choice of the spin projection of massive particles
\footnote{ In order to prove that the formula appeared in \citep{Boels:2011zz} is same as the above relation for a specific choice of the spin projection of massive vectors, 
 we use the following decomposition of little group covariant massive spinor-helicity variables \citep{Arkani-Hamed:2017jhn}
$ \lambda ^\alpha _I=\lambda ^\alpha \xi ^+_I -\eta ^\alpha \xi ^-_I\,,$
where $\lambda _\alpha ,\eta _\alpha$ are massless spinor-helicity variables and satisfy $\langle \lambda \eta \rangle =m$ and $\xi ^{\pm I}$ are suitable SU(2) basis vectors.  Setting the particle with momentum $p_1$ with $s_z=+1$ and particle with momentum $p_n$ with $s_z=-1$ in the amplitude,  we find that
$
\langle \textbf{1}\textbf{n}\rangle _{(+,-)} \longrightarrow  \langle \eta _1 \lambda_n\rangle \,.$
Therefore,  we can recast the relation \eqref{vector-scalar} with the massive particles are being in this specific spin state as follows
\begin{align*}
\mathcal{A}_{n}[\textbf{1}_+,2^+,\ldots, \textbf{n}_-]=\left(\frac{\langle \eta _1 \lambda_n\rangle}{\langle \lambda_1 \eta _1 \rangle}\right)^2  \mathcal{A}_{n}[\textbf{1}^0,2^+,\ldots, \textbf{n}^0]\,. 
\end{align*}
This is the relation that appeared in \cite{Boels:2011zz}. }. Furthermore,   using the expressions for massive amplitudes in \citep{Ballav:2020ese}, we have explicitly verified this covariant formula in case of lower-point amplitudes, such as four- and five- particle amplitudes. 

Now the $n$-point amplitude with a pair massive scalars and $(n-2)$ positive helicity gluons is already known \cite{Ferrario:2006np}:  
\begin{align}
\mathcal{A}_n[\textbf{1}^0,2^+,\cdots,(n-1)^+,\textbf{n}^0]=g^{n-2}\tfrac{ m^2[2|\prod _{k=3}^{n-2}\left((s_{1\ldots k}-m^2)-\slashed{p}_k \cdot \slashed{p}_{1,k-1}\right)|n-1]}{(s_{12}-m^2)(s_{123}-m^2)\cdots (s_{12\ldots (n-2)}-m^2)\langle 23 \rangle \langle 34\rangle \cdots \langle (n-2)(n-1) \rangle}\,,
\label{nptscalar1}
\end{align}
where the Mandelstam variables and $p_{1,l}$ are defined as follows 
\begin{align}
s_{1\ldots l}:=\left(p_1+\cdots+p_l\right)^2\,,\qquad p_{1,l}:=p_1+\cdots+p_l\,.
\end{align}
In equation \eqref{nptscalar1}, we have introduced short hand notation for spinor products defined as follows 
\begin{align}
[a|\slashed{p}_i\cdot \slashed{p}_j|b]=\tilde{\lambda} _{a\dot{\alpha}}p_i^{\dot{\alpha}\alpha}p_{j\alpha \dot{\beta}}\tilde{\lambda}^{\dot{\beta}}_b\,.
\end{align}
Note that we treat the momentum product $\slashed{p}_i \cdot \slashed{p}_j$ as SU(2) matrix valued product $p_i^{\dot{\alpha}\alpha}p_{j\alpha \dot{\beta}}$ when being contracted with spinor helicity variables.  We follow this notation throughout this paper.  The product appearing in the numerator of the formula \eqref{nptscalar1} is defined as 
\begin{align}
 [2|\prod _{k=3}^{n-2}\mathcal{B}_k|n-1]
& :=[2|\mathcal{B}_3 \cdot \mathcal{B}_4\cdot \ldots \cdot \mathcal{B}_{n-2} |n-1]
\end{align}

Substituting the scalar amplitude in \eqref{vector-scalar},  we therefore find the following simple expression for the $n$-point amplitude with a pair massive vector bosons and $(n-2)$ positive helicity gluons (for $n>3$):{\footnote{For massive particles with spin-$s$ and all positive helicity gluons, the formula of the amplitude has recently appeared in \cite{Lazopoulos:2021mna}.}}
\begin{align}
\mathcal{A}_n[\textbf{1},2^+,\cdots,(n-1)^+,\textbf{n}]=g^{n-2}\tfrac{\langle \textbf{1}\textbf{n}\rangle ^2 [2|\prod _{k=3}^{n-2}\left((s_{1\ldots k}-m^2)-\slashed{p}_k \cdot \slashed{p}_{1,k-1}\right)|n-1]}{(s_{12}-m^2)(s_{123}-m^2)\cdots (s_{12\ldots (n-2)}-m^2)\langle 23 \rangle \langle 34\rangle \cdots \langle (n-2)(n-1) \rangle}\,. 
\label{nptvector}
\end{align}

\subsection{Inductive proof using covariant recursion}\label{section-3.1}
In this section, we present an inductive proof of the above formula in \eqref{nptvector} using the covariant recursion that was reviewed in Section \ref{mmshift}.
To set up the induction, we first of all have to ensure that the four- and five-point amplitudes that have been calculated previously in \cite{Ballav:2020ese} are consistent with the general expression. We perform this check in Appendix \ref{appendix: a}. 

\begin{figure}
\includegraphics[scale=.37]{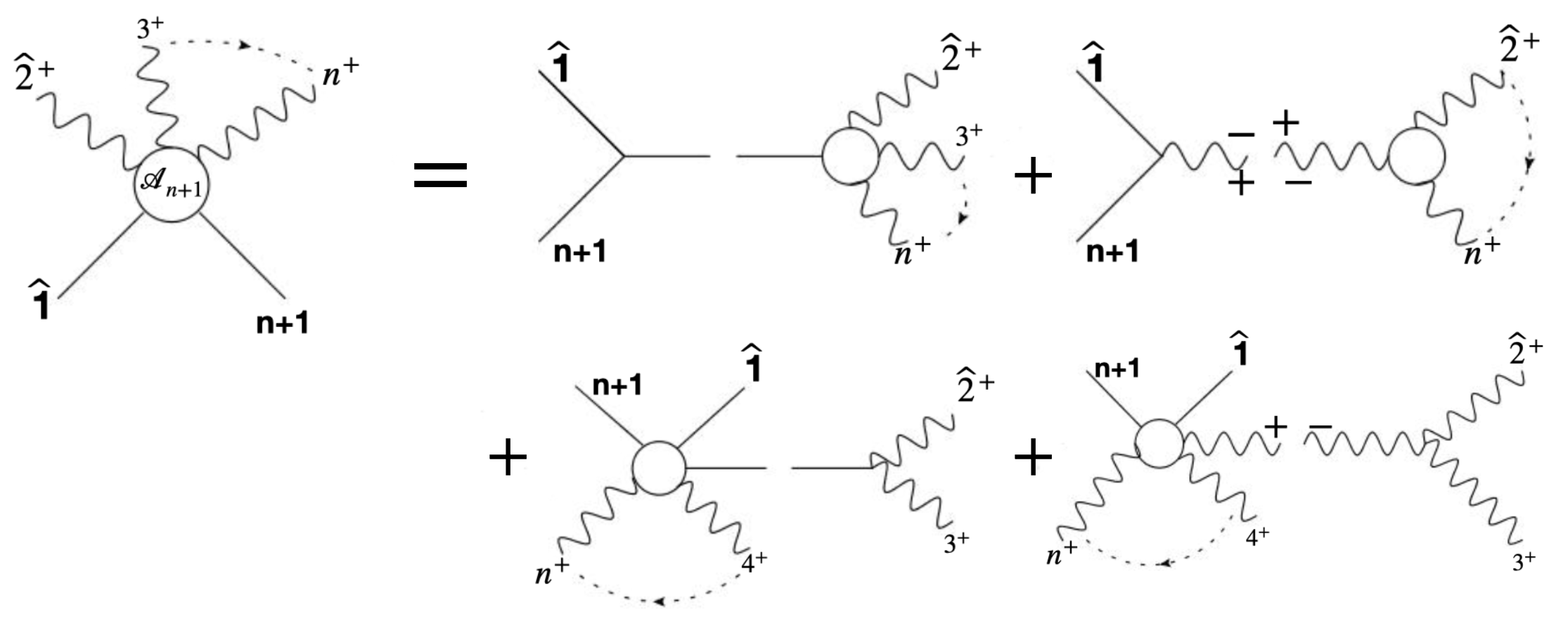}
\caption{Pictorial representation of covariant recursion with $[\textbf{1}2^+\rangle$ shift. }
\label{allplusdiag}
\end{figure}

Given the match of the lower-point amplitudes we now assume that the expression \eqref{nptvector} is true for $n$-particle amplitude and use this to construct $(n+1)$-particle amplitude. We use the $[\textbf{1}2^+\rangle$ shift which corresponds to the shifts of the following spinor-helicity variables:
\begin{align}
|\widehat{1}^I]=|1^I]-\frac{z}{m}[1^I2]|2]\,,\qquad |\widehat{2}\rangle=|2\rangle +\frac{z}{m}p_1|2]\,,  \label{covriantshiftsec3}
\end{align}
whereas the spinor-helicity variables $|1^I\rangle $ and $|2]$ remain unchanged.  With this particular shift, all possible channels that contribute to the $\mathcal{A}_{n+1}$ amplitude are shown in Figure \ref{allplusdiag}.
The first three diagrams do not contribute to the amplitude due to following reasons: a) the contribution from the first diagram vanishes due to the vanishing of the right subamplitude involving a single massive vector boson,  b) the contribution from the second diagram vanishes due to the vanishing of the pure gluon subamplitude with either all positive helicity gluons or a single negative helicity gluon,  c) the contribution from the third diagram vanishes because a massive vector boson cannot decay into two identical gluons.  Thus we only have to compute the contribution from the fourth diagram.  

From the only non-vanishing diagram,  we get a simple pole in the $z$-plane by setting the shifted propagator $\widehat{s}_{23}$ on-shell
\begin{align}
(\widehat{p}_2+p_3)^2=0 \Rightarrow z_I=\frac{m \langle 23\rangle}{\langle 3|p_1|2]}\,.
\end{align}
The $(n+1)$-particle amplitude $\mathcal{A}_{n+1}\left[\textbf{1},2^+,\cdots ,n^+,\textbf{n+1}\right]$ is therefore assembled from the $n$-point  and 3-point subamplitudes
\begin{align}
\mathcal{A}_{n+1}= \mathcal{A}_{n}[\widehat{\textbf{1}},\widehat{I}^+,4^+,\cdots,n^+,(\textbf{n+1})] \frac{1}{s_{23}} \mathcal{A}_3[\widehat{I}^-,\widehat{2}^+,3^+]\,.
\end{align}
Here we abbreviate $\mathcal{A}_{n+1}\left[\textbf{1},2^+,\cdots ,n^+,\textbf{n+1}\right] $ as $\mathcal{A}_{n+1}$.  The alternative helicity configuration of the internal states does not contribute to the amplitude due to the vanishing of all-positive-helicity three-particle gluon amplitude.  Using the expression for $n$-point amplitude in equation \eqref{nptvector}, we get the left subamplitude but now with shifted momenta
\begin{align}
\mathcal{A}_{n}[\widehat{\textbf{1}}^0,\widehat{I}^+,4^+,\cdots,n^+,(\textbf{n+1})^0]=g^{n-2}\tfrac{\langle \textbf{1}(\textbf{n+1})\rangle ^2[\widehat{I}|\prod _{k=4}^{n-1}\left((\widehat{S}_{1I \ldots k}-m^2)-\slashed{p}_k \cdot \widehat{\slashed{P}}_{1,k-1}\right)|n]}{(\widehat{S}_{1I}-m^2)(\widehat{S}_{1I4}-m^2)\cdots (\widehat{S}_{1I\ldots (n-1)}-m^2)\langle \widehat{I}4 \rangle \langle 45\rangle \cdots \langle (n-1)n \rangle}\,. \label{n+1scalar}
\end{align}
Here $\widehat{S}$ $(\widehat{P})$ are the Mandelstam (momentum) variable with the shifted momenta
\begin{align}
\widehat{S}_{1I \ldots r}=(\widehat{p}_1+\widehat{p}_I+p_4+\cdots +p_r)^2\,,\quad \widehat{P}_{1,r}=(\widehat{p}_1+\widehat{p}_I+\cdots +p_r)\,.
\end{align}
The internal momentum $\widehat{p}_I$ in this channel is $\widehat{p}_2+p_3$.  Therefore we find that these shifted variables can be simply expressed in terms of the unshifted variables as
\begin{align*}
\widehat{S}_{1 \ldots r}=(\widehat{p}_1+\widehat{p}_2+p_3+\cdots +p_r)^2=s_{1\ldots r}\,,\quad \widehat{P}_{1,r}=(\widehat{p}_1+\widehat{p}_2+p_3+\cdots +p_r)=p_{1,r}\,.
\end{align*}
Using these simplifications and gluing the three-particle gluon amplitude along with the unshifted propagator $\frac{1}{s_{23}}$ onto the left subamplitude,  we obtain
\begin{align}
\mathcal{A}_{n+1}=\tfrac{g^{n-1}\langle \textbf{1}(\textbf{n+1})\rangle ^2[\widehat{I}|\prod _{k=4}^{n-1}\left((s_{1\ldots k}-m^2)-\slashed{p}_k \cdot \slashed{p}_{1,k-1}\right)|n]}{(s_{123}-m^2)(s_{1..4}-m^2)\cdots (s_{1\ldots (n-1)}-m^2)\langle \widehat{I}4 \rangle \langle 45\rangle \cdots \langle (n-1)n \rangle} \times \tfrac{[23]^2}{\langle 23\rangle [\widehat{I}2][\widehat{I}3]}
\end{align}
It remains to simplify the terms with the shifted massless spinor-helicity variable $\widehat{I}$ associated with the momentum of the exchange particle.  We collect all such terms and rewrite them as 
\begin{align}
\tfrac{[\widehat{I}|\prod _{k=4}^{n-1}\left((s_{1\ldots k}-m^2)-\slashed{p}_k \cdot \slashed{p}_{1,k-1}\right)|n]}{\langle \widehat{I}4\rangle [\widehat{I}2][\widehat{I}3]}
&= \tfrac{[2|p_1|\widehat{I}\rangle  [\widehat{I}|\prod _{k=4}^{n-1}\left((s_{1\ldots k}-m^2)-\slashed{p}_k \cdot \slashed{p}_{1,k-1}\right)|n]}{\langle 4|p_3|2][21_I]\langle 1^I\widehat{I}\rangle[\widehat{I}3]}\cr
&= \tfrac{[2|p_1\cdot (p_2+p_3)\prod _{k=4}^{n-1}\left((s_{1\ldots k}-m^2)-\slashed{p}_k \cdot \slashed{p}_{1,k-1}\right)|n]}{[23]^2\langle 34\rangle (s_{12}-m^2)}
\end{align}
We have replaced $\widehat{p}_2 \rightarrow p_2$ in the intermediate step while multiplying with $\langle \widehat{I}|p_1|2]$.  This is allowed because
\begin{align}
\langle 1^I\widehat{2}\rangle=\langle 1^I2\rangle -z_I[1^I2]\Rightarrow \langle 1^I\widehat{2}\rangle [1_I2]=\langle 1^I2\rangle [1_I2]\,,
\end{align}
where we have used $[1^I2][1_I2]=-m[22]=0$.  Using the following identity
\begin{align}
[2|\slashed{p}_1 \cdot (\slashed{p}_{2}+\slashed{p}_3)=[2|\left\lbrace(s_{123}-m^2)-\slashed{p}_3\cdot (\slashed{p}_1+\slashed{p}_2)\right\rbrace \,,
\end{align}
we finally obtain the $(n+1)$-point amplitude in the form
\begin{equation}
\mathcal{A}_{n+1}[\textbf{1},2^+,\ldots,n^+, \textbf{n+1}]=\frac{g^{n-1} \langle \textbf{1}(\textbf{n+1})\rangle ^2[2|\prod _{k=3}^{n-1}\left((s_{1\ldots k}-m^2)-\slashed{p}_k \cdot \slashed{p}_{1,k-1}\right)|n]}{(s_{12}-m^2)(s_{123}-m^2)\cdots (s_{12\ldots (n-1)}-m^2)\langle 23 \rangle \langle 34\rangle \cdots \langle (n-1)n \rangle}\,.  \label{vectoramp}
\end{equation}
This completes the inductive proof of $n$-particle amplitude with all plus helicity gluons and a pair of massive vector bosons.  The scattering amplitude with two massive vector bosons and all \textit{minus} helicity gluons can be read off from the expression in \eqref{nptvector} by replacing all the angle brackets with square brackets and vice-versa
\begin{equation}
\mathcal{A}_{n}[\textbf{1},2^-,\ldots,(n-1)^-, \textbf{n}]=g^{n-2}\,\tfrac{ [ \textbf{1}\textbf{n}]^2\langle 2|\prod _{k=3}^{n-2}\left((s_{1\ldots k}-m^2)-\slashed{p}_k \cdot \slashed{p}_{1,k-1}\right)|(n-1)\rangle }{(s_{12}-m^2)(s_{123}-m^2)\cdots (s_{12\ldots (n-2)}-m^2) [23][34] \cdots [(n-2)(n-1)]}\,. 
\end{equation}
It is instructive to check the high energy limit of the massive vector boson amplitude \eqref{nptvector}.  Due to the presence of angle bracket $\langle \textbf{1}\textbf{n}\rangle ^2$,  the only non-zero contribution comes from the component of the massive amplitude with both massive particles having negative helicity in the high energy limit \citep{Arkani-Hamed:2017jhn}. 

\subsection{Matching the MHV amplitude in the high energy limit}\label{masslessMHVchk} 
In this section,  we recover the known massless amplitude from the massive vector boson amplitude with all positive helicity gluons.  We show that the finite energy amplitude in equation \eqref{nptvector} reproduces correct MHV amplitude in the high energy limit for negative helicity configuration of massive particles in this limit. The massless amplitude is given by
\begin{align}
\mathcal{A}_{n}&[1^-,2^+,\ldots, (n-1)^+,n^-]=g^{n-2}\frac{\langle 1n \rangle ^2[2|\prod _{k=3}^{n-2}\left(s_{1\ldots k}-\slashed{p}_k \cdot \slashed{p}_{1,k-1}\right)|(n-1)]}{s_{12}s_{123}\cdots s_{12\ldots (n-2)}\langle 23 \rangle \langle 34\rangle \cdots \langle (n-2)(n-1) \rangle}\\
&=g^{n-2}\frac{\langle 1n\rangle ^3}{\langle 12 \rangle \langle 23\rangle \cdots \langle (n-1)n\rangle} \frac{[2|\prod _{k=3}^{n-2}\left(s_{1\ldots k}-\slashed{p}_k \cdot \slashed{p}_{1,k-1}\right)|(n-1)]\langle (n-1)n\rangle}{[21]s_{123}\cdots s_{12\ldots (n-2)} \langle 1n\rangle} \,.
\end{align}   
Let us consider the non-trivial part of this amplitude
\begin{align}
\mathcal{M}_{n}:= \frac{[2|\prod _{k=3}^{n-2}\left(s_{1\ldots k}-\slashed{p}_k \cdot \slashed{p}_{1,k-1}\right)\cdot \slashed{p}_{n-1}| n\rangle}{[21]s_{123}\cdots s_{12\ldots (n-2)} \langle 1n\rangle} \,. \label{mn+1}
\end{align}
We simplify the product in the numerator by using $n$-th massless particles momentum conservation and identity \eqref{Pauliid} repeatedly.  We start with the $k=n-2$ term and use momentum conservation to write\footnote{For a single $SU(2)$ matrix valued momentum variable contracted to spinor-helicity variable, we omit the slash notation as in standard literature: $\slashed{p}_i| j\rangle \equiv p_i |j\rangle$.}
\begin{align}
\left(s_{1\ldots n-2}-\slashed{p}_{n-2} \cdot \slashed{p}_{1,n-3}\right)\cdot \slashed{p}_{n-1}|n\rangle =s_{1\ldots n-2}\,p_{n-1}|n\rangle+\slashed{p}_{n-2} \cdot \slashed{p}_{n}\cdot \slashed{p}_{n-1}|n\rangle\,. \label{masslessid1}
\end{align}
Let us explain the notation we are using here for generic momenta and spinor-helicity variables
\begin{align}
(s_{ij}-\slashed{p}_l\cdot \slashed{p}_m)|r\rangle \equiv s_{ij}\lambda _{r\alpha} -p_{l\alpha \dot{\alpha}}p_m^{\dot{\alpha}\beta}\lambda _{r\beta}\,.
\end{align}
Here the Greek indices are the SL(2,$\mathbb{C}$) Lorentz indices.  Going back to \eqref{masslessid1}, we use \eqref{Pauliid} to express the second term as follows
\begin{align}
\slashed{p}_{n-2} \cdot \slashed{p}_{n}\cdot \slashed{p}_{n-1}|n\rangle =\left(2p_{n-1} \cdot p_{n}\right)p_{n-2}|n\rangle  \,.
\end{align}
Here we use the fact that $p_{n}|n\rangle =0$.  Incorporating this with \eqref{masslessid1}, we obtain
\begin{align}
\left(s_{1\ldots n-2}-\slashed{p}_{n-2} \cdot \slashed{p}_{1,n-3}\right)\cdot \slashed{p}_{n-1}|n\rangle = s_{1\ldots n-2}(p_{n-2} +p_{n-1})|n\rangle \,.
\end{align}
Now we include the next term in the product in \eqref{mn+1} and use the above result to write
\begin{align}
\prod _{k=n-3}^{n-2}\left(s_{1\ldots k}-\slashed{p}_k \cdot \slashed{p}_{1,k-1}\right)\cdot \slashed{p}_{n-1} |n\rangle =s_{1\ldots n-2}s_{1\ldots n-3}(p_{n-2} +p_{n-1})|n\rangle \cr
 -\slashed{p}_{n-3} \cdot \slashed{p}_{1,n-4}\cdot (\slashed{p}_{n-2} +\slashed{p}_{n-1})|n\rangle\,.
\end{align}
We can again simplify the second term using momentum conservation and \eqref{Pauliid} to get
\begin{align}
-\slashed{p}_{n-3} \cdot \slashed{p}_{1,n-4}\cdot (\slashed{p}_{n-2} +\slashed{p}_{n-1})|n\rangle =\left(s_{1\ldots n-3} s_{1\ldots n-2}\right) p_{n-3}|n\rangle \,.
\end{align}
Hence we derive
\begin{align}
\prod _{k=n-3}^{n-2}\left(s_{1\ldots k}-\slashed{p}_k \cdot \slashed{p}_{1,k-1}\right)\cdot \slashed{p}_{n-1} |n\rangle =s_{1\ldots n-2}s_{1\ldots n-3}(p_{n-3}+p_{n-2} +p_{n-1})|n\rangle \,.
\end{align}
This trend continues to follow and we obtain the following identity
\begin{align}
\prod _{k=3}^{n-2}\left(s_{1\ldots k}-\slashed{p}_k \cdot \slashed{p}_{1,k-1}\right)\cdot \slashed{p}_{n-1} |n\rangle =\prod _{k=3}^{n-2} s_{1\ldots k} (p_3+\cdots +p_{n-2}+p_{n-1})|n\rangle\,. \label{masslessid}
\end{align}
Therefore we have established that $\mathcal{M}_{n}=-1$.  Thus the only non-vanishing high energy limit of the massive vector boson amplitude \eqref{vectoramp} reproduces MHV amplitude
\begin{align}
\mathcal{A}_{n}&[1^-,2^+,\ldots, (n-1)^+,n^-]=-g^{n-2}\frac{\langle 1n\rangle ^3}{\langle 12 \rangle \langle 23\rangle \cdots \langle (n-1)n\rangle}\,.
\end{align} 
This provides a primary consistency check for the massive $n$-point amplitude in \eqref{nptvector}.

Having shown that the covariant recursion relations can be used to inductively prove the formula \eqref{nptvector} of the massive analogue of MHV amplitude,  it is worthwhile to mention that one could do the same by using the BCFW recursion relations as well. However, the real benefit of the covariant  recursion will be apparent in the next section where we will show that a similar application of covariant recursion can be achieved in the case in which one of the gluon has flipped helicity.

\section{Scattering of massive vector bosons with a flipped helicity gluon }\label{section-4}

\begin{figure}
\includegraphics[scale=.38]{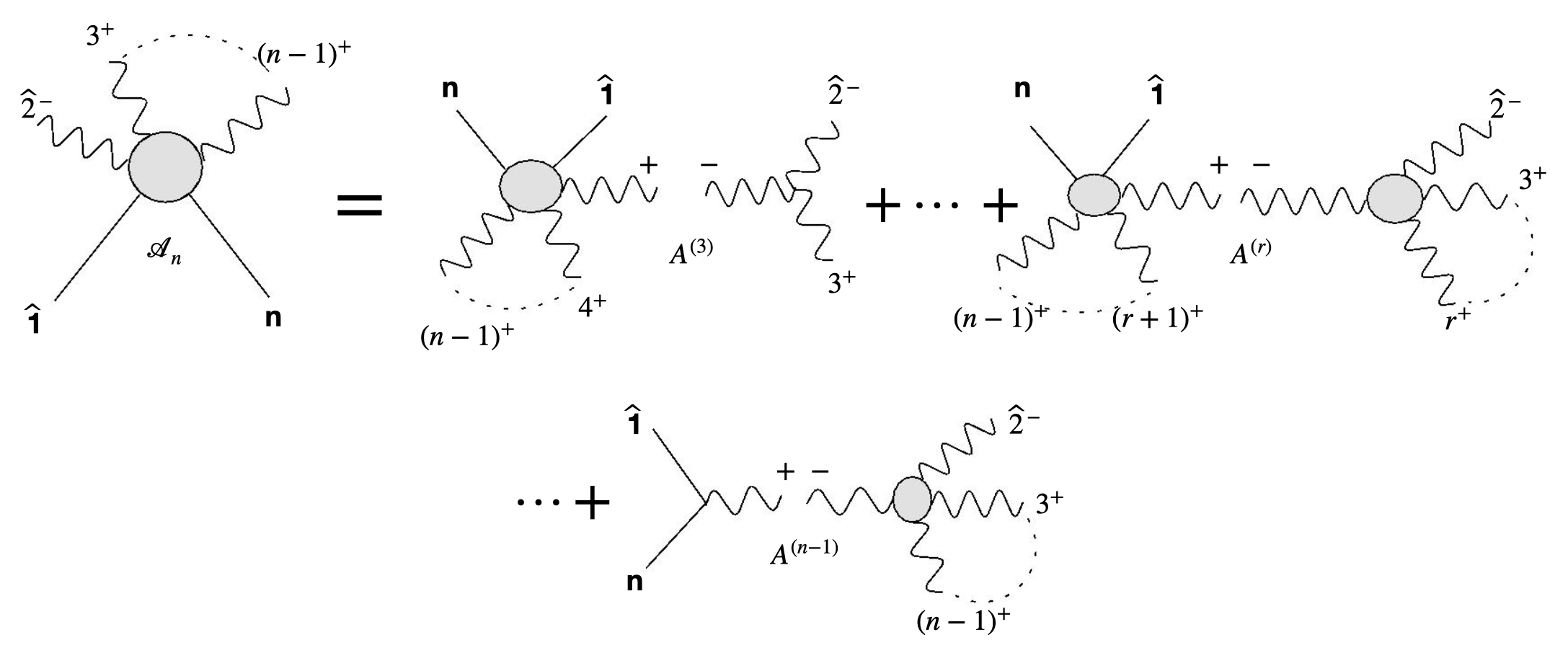}
\caption{Pictorial representation of covariant recursion with $[2^-\textbf{1} \rangle$ shift for $\mathcal{A}_n[\textbf{1},2^-,3^+,\ldots,\textbf{n}]$.}
\label{flippedhelicitydiag}
\end{figure}

We consider tree-level colour-ordered amplitude involving a pair of massive vector bosons, one minus helicity gluon and arbitrary number of positive helicity gluons.  For simplicity, we assume that the massive particles and the negative helicity gluon are colour adjacent to each other,  as indicated in the Figure \ref{flippedhelicitydiag}. We shall see later that the scattering amplitude with this specific external particle configuration leads to the NMHV pure gluon amplitude in the high energy limit.

If one had used the usual BCFW shift to compute this amplitude, one would end up with subamplitudes involving the same configuration as the one we set out to compute (i.e. involving two massive vector bosons and helicity flipped gluons).  In the absence of an ansatz one would need to use the recursion relation iteratively to compute those subamplitudes that appear in a given recursion. This would make the computation technically involved.

Instead, we use the the massive-massless shift $[2^-\textbf{1}\rangle$ of the type $[-\textbf{m}\rangle$ to compute this particular $n$-point amplitude. This shift corresponds to the following deformation in terms of the spinor-helicity variables :
\begin{align}
|\widehat{2}]=|2]+\frac{z}{m}p_1|2\rangle \,, \qquad |\widehat{1}^I\rangle =|1^I\rangle -\frac{z}{m}\langle 21^I\rangle |2\rangle \,. \label{21shift}
\end{align}  
With this shift and the chosen configuration of external particles, the different scattering channels that contribute to the amplitude in the covariant recursion are shown in Figure \ref{flippedhelicitydiag}. As one can see, all the relevant subamplitudes have already been computed: either they involve only pure gluon amplitudes or they involve two massiee vector bosons and all positive helicity gluons. 

As we are considering only minimal coupling while computing the amplitudes,  the exchange particles can be either massive vector boson or gluon.  But the exchange particle can not be a massive vector boson as a massive spin-1 particle can not decay into two massless gluons.  Due to the $[2^-\textbf{1}\rangle$ shift, particles with momenta $\widehat{p}_1$ and $\widehat{p}_2$ are always attached to different subamplitudes in the   diagramatic expansion of the colour-ordered amplitude.  Now the subamplitude involving the external momentum $\widehat{p}_2$ will always have only positive helicity gluons as external states.  But such pure gluon amplitudes with at most one opposite helicity vanishes, except for the three-particle amplitude.  Hence the internal state attached to this subamplitude must be a negative helicity gluon.  Again,  due to the choice of the massive-massless shift $[2^-\textbf{1}\rangle$,  the first diagram in Figure \ref{flippedhelicitydiag} is non-vanishing only for the helicity configuration indicated.  

The $n$-particle amplitude obtained by summing over the various diagrams can thus be written as follows
\begin{align}
\mathcal{A}_n[\textbf{1},2^-,3^+,\ldots, (n-1)^+,\textbf{n}]&=\sum _{r=3}^{n-1}\mathcal{A}_L[\widehat{\textbf{1}},\widehat{I}^+,(r+1)^+,\ldots,\textbf{n}] \frac{1}{s_{23\ldots r}}\mathcal{A}_R[\widehat{I}^-,\widehat{2}^-,3^+,\ldots,r^+] \,,\label{npointrecursion}
\end{align}
where $s_{2\ldots r}=(\sum _{i=2}^{r}p_i)^2$.  Here the subamplitudes are on-shell; that is, they are functions of shifted momenta and spinor-helicity variables. The right subamplitude is a pure-gluon amplitude and is given by the Parke-Taylor formula
\begin{align}
\mathcal{A}_R[\widehat{I}^-,\widehat{2}^-,3^+,\ldots,r^+]=g^{r-2}\frac{\langle \widehat{I}2\rangle ^3}{\langle 23\rangle \langle 34\rangle \ldots \langle r\widehat{I}\rangle}\,.\label{puregluonflipped}
\end{align}
The left subamplitude involving two massive vector bosons and all positive helicity gluons is known from the previous section and is given by (see equation \eqref{nptvector})
\begin{align}
\mathcal{A}_L[\widehat{\textbf{1}},\widehat{I}^+,(r+1)^+,\ldots,\textbf{n}]= g^{n-r}\tfrac{\langle \widehat{\textbf{1}}\textbf{n}\rangle ^2[\widehat{I}|\prod _{k=r+1}^{n-2}\left\lbrace (\widehat{S}_{1I \cdots k}-m^2)-\slashed{p}_k\cdot \widehat{\slashed{P}}_{1,k-1}\right\rbrace|n-1]}{(\widehat{S}_{1I}-m^2)\ldots (\widehat{S}_{1I(r+1)\ldots (n-2)}-m^2)\langle \widehat{I}(r+1)\rangle \ldots \langle (n-2)(n-1)\rangle}\,. \label{leftsubflipped}
\end{align} 
This takes care of all but one diagram that appears in the covariant recursion. The last diagram in Figure \ref{flippedhelicitydiag} (which corresponds to $r=n-1$), has to be treated separately and we shall come to the evaluation of this diagram towards the end of this section.  

Let us now simplify the expression in \eqref{leftsubflipped} and write it purely in terms of the external momenta. Using $\widehat{p_I}=\widehat{p}_2 +\sum _{i=3}^r p_i$,  the shifted Mandelstam variables ($\widehat{S}$) and momenta $(\widehat{P})$ can be expressed (for $ k \in \{ r+1, \ldots (n-2)\}$) as follows
\begin{align}
&\widehat{S}_{1\ldots k}=(\widehat{p}_1+\widehat{p}_I+\cdots + p_k)^2=(p_1+p_2+\cdots + p_k)^2=s_{1\ldots k} \, ,  \cr
&\widehat{P}_{1,k-1}=(\widehat{p}_1+\widehat{p}_I+\cdots +p_{k-1})=(p_1+p_2+\cdots + p_{k-1})=p_{1,k-1}.
\end{align}
Substituting these into the left subamplitude \eqref{leftsubflipped} and then gluing this with the pure gluon amplitude \eqref{puregluonflipped} and taking care of the unshifted propagator $\frac{1}{s_{2\ldots r}}$, we get the contribution to the $n$-particle amplitude from the $r$-th term in the covariant recursion \eqref{npointrecursion}:
\begin{align}
A^{(r)}:=g^{n-2}\tfrac{\langle \widehat{\textbf{1}}\textbf{n}\rangle^2\langle \widehat{I}2\rangle ^3[\widehat{I}|\prod _{k=r+1}^{n-2}\left\lbrace (s_{1\cdots k}-m^2)-\slashed{p}_k\cdot \slashed{p}_{1,k-1}\right\rbrace|n-1]}{s_{23\ldots r}(s_{12\ldots r}-m^2)\ldots (s_{12..r(r+1)\ldots (n-2)}-m^2)  \langle 23\rangle \langle 34\rangle \ldots \langle r\widehat{I}\rangle   \langle \widehat{I}(r+1)\rangle \ldots \langle (n-2)(n-1)\rangle }\,. \label{arshiftedflipped}
\end{align}
Here $r \in \{3, 4, \ldots (n-2)\}$. 
We would like to note that the product of angle brackets in the denominator, involving the massless spinor-helicity variables do not include $\langle r (r+1)\rangle$ bracket as the $r$- and $(r+1)$-th massless external legs do not attach to same subamplitude.  Next we express all the spinor products in $A^{(r)}$ involving the intermediate spinor-helicity variable $|\widehat{I}\rangle$ in terms of the spinor-helicity variables of the external particles.  In order to do that, we collect all such terms as
\begin{align}
\chi _{r,I} =\frac{\langle \widehat{I}2\rangle ^3[\widehat{I}|\prod _{k=r+1}^{n-2}\left\lbrace (s_{1\cdots k}-m^2)-\slashed{p}_k\cdot \slashed{p}_{1,k-1}\right\rbrace|n-1]}{\langle \widehat{I}(r+1)\rangle \langle r\widehat{I}\rangle} \,.
\end{align}
We use the following identities 
\begin{align*}
\langle 2\widehat{I}\rangle [\widehat{I}|\mathcal{B} |n-1]&=\langle 2|\slashed{p}_{3,r}\cdot \mathcal{B} |n-1] \,, \quad  \mathcal{B}=\prod _{k=r+1}^{n-2}\left\lbrace (s_{1\cdots k}-m^2)-\slashed{p}_k\cdot \slashed{p}_{1,k-1}\right\rbrace\,,\\
\frac{\langle \widehat{I}2\rangle}{ \langle \widehat{I} r\rangle}&=\frac{\langle 2|\slashed{p}_1.\slashed{p}_{3,r}|2\rangle}{ \langle 2|\slashed{p}_1.\slashed{p}_{2,r-1}|r\rangle}\,, \quad
\frac{\langle \widehat{I}2\rangle}{ \langle \widehat{I} (r+1)\rangle}=\frac{\langle 2|\slashed{p}_1.\slashed{p}_{3,r}|2\rangle}{ \langle 2|\slashed{p}_1.\slashed{p}_{2,r}|r+1\rangle}\,,
\end{align*}
to write
\begin{align}
\chi _{r,I} =\frac{\langle 2|\slashed{p}_1.\slashed{p}_{3,r}|2\rangle^2\langle 2|\slashed{p}_{3,r}\prod _{k=r+1}^{n-2}\left\lbrace (s_{1\cdots k}-m^2)-\slashed{p}_k\cdot \slashed{p}_{1,k-1}\right\rbrace|n-1]}{\langle 2|\slashed{p}_1.\slashed{p}_{2,r-1}|r\rangle\langle 2|\slashed{p}_1.\slashed{p}_{2,r}|r+1\rangle} \,. \label{idependflipped}
\end{align}

It only remains to evaluate the shifted spinor product $\langle \widehat{\textbf{1}}\textbf{n}\rangle$. The simple pole, associated with scattering channels (except $s_{1n}$) in Figure \ref{flippedhelicitydiag} is obtained by setting the shifted propagator $\widehat{s}_{2\cdots r}$ on-shell: 
\begin{align}
(p_2+z_{(r)}q+p_3 +\ldots +p_r)^2=0 \Rightarrow z_{(r)}=-\,\frac{m p_{2,r}^2}{\langle 2|\slashed{p}_1\cdot \slashed{p}_{3,r}|2\rangle}\,.
\end{align}
We then use the definition of the shifted massive spinor-helicity variable in \eqref{21shift} at $z=z_{(r)}$ to express the spinor product $\langle \widehat{1}^In^J\rangle$ as 
\begin{align}
\langle \widehat{1}^In^J\rangle =\langle 1^In^J\rangle +\frac{p_{2,r}^2}{\langle 2|\slashed{p}_1\cdot \slashed{p}_{3,r}|2\rangle} \langle 1^I2\rangle \langle 2n^J\rangle \,.  \label{shiftedspinor1nflipped}
\end{align}
Substituting the expressions \eqref{idependflipped} and \eqref{shiftedspinor1nflipped} in \eqref{arshiftedflipped}, one can finally rewrite $A^{(r)}$  in terms of the on-shell external variables: 
\begin{align}
A^{(r)}=g^{n-2}\tfrac{\langle 2|\slashed{p}_{3,r}\cdot \prod _{k=r+1}^{n-2}\left\lbrace (s_{1\cdots k}-m^2)-\slashed{p}_k\cdot \slashed{p}_{1,k-1}\right\rbrace|n-1]\left( \langle 2|\slashed{p}_1.\slashed{p}_{3,r}|2\rangle \langle \textbf{1}\textbf{n}\rangle +p_{2,r}^2\langle \textbf{1}2\rangle \langle 2\textbf{n}\rangle \right)^2 \langle r(r+1)\rangle}{s_{23\ldots r}(s_{12\ldots r}-m^2)\ldots (s_{12\ldots (n-2)}-m^2) \langle 23\rangle \langle 34\rangle \ldots  \langle (n-2)(n-1)\rangle\langle 2|\slashed{p}_1.\slashed{p}_{2,r-1}|r\rangle\langle 2|\slashed{p}_1.\slashed{p}_{2,r}|r+1\rangle}\,. \label{archannel}
\end{align}

We now analyze the last diagram in Figure \ref{flippedhelicitydiag}, which corresponds to the $r=n-1$ term  in the covariant recursion. We have to treat this term separately because the left subamplitude involving two massive vector bosons and a single positive helicity gluon cannot be read off from the formula \eqref{nptvector} (we explicitly assumed $n>3$ in that derivation). Instead we simply glue the three-particle amplitude in \eqref{minimalthreepoint} for positive helicity gluon along with the pure gluon amplitude \eqref{puregluonflipped} for $r=n-1$ and the unshifted propagator $\frac{1}{s_{1n}}$
\begin{align}
A^{(n-1)} =g^{n-2} \frac{-\,\langle \widehat{\textbf{1}}\textbf{n}\rangle ^2 \langle \widehat{I}2\rangle ^2 s_{3,(n-1)}}{s_{1n}\langle 23\rangle \langle 34\rangle \cdots \langle (n-2)(n-1)\rangle \langle \widehat{I}|p_n|\widehat{2}]\langle (n-1)\widehat{I}\rangle} \,.\label{an-1}
\end{align}
Again, we have to simplify the terms with $|\widehat{I}\rangle$ and evaluate the shifted spinor products,  obtained by setting the shifted propagator $\widehat{s}_{1n}$ on-shell:
\begin{align}
z_{(n-1)}=\frac{m(p_1+p_n)^2}{\langle 2|\slashed{p}_1\cdot \slashed{p}_n)|2 \rangle }\,.  \label{poleforn-1}
\end{align}
 Firstly,  by noting the following identities
\begin{align}
\langle 2|\slashed{p}_1\cdot \widehat{\slashed{p}}_I\cdot \slashed{p}_n|\widehat{2}] &=m^2 \langle 2|(p_1+p_n)|\widehat{2}]  \,,\\
\langle 2|\slashed{p}_1\cdot (\widehat{\slashed{p}}_1+\slashed{p}_n)|n-1\rangle  &= \langle 2|\slashed{p}_1\cdot \slashed{p}_n|n-1\rangle +m^2\langle 2(n-1)\rangle \,,
\end{align}
we get rid of the internal momentum dependence of $A^{(n-1)}$ as follows
\eqst{
\frac{-\,\langle \widehat{I}2\rangle ^2}{\langle \widehat{I}|p_n|\widehat{2}]\langle (n-1) \widehat{I}\rangle}=\frac{\langle 2| \slashed{p}_1 \cdot \slashed{p}_n |2\rangle ^2}{\langle 2|\slashed{p}_1\cdot \widehat{\slashed{p}}_I\cdot \slashed{p}_n|\widehat{2}] \langle 2|\slashed{p}_1\cdot (\widehat{\slashed{p}}_1+\slashed{p}_n)|n-1\rangle } \cr
=\frac{\langle 2| \slashed{p}_1 \cdot \slashed{p}_n |2\rangle ^2}{m^2\left( \langle 2|p_1|\widehat{2}] + \langle 2|p_n|\widehat{2}] \right) \left( \langle 2|\slashed{p}_1\cdot \slashed{p}_n|n-1\rangle +m^2\langle 2(n-1)\rangle \right)}\,.}
Secondly, we calculate the shifted spinor products appearing in this expression and in \eqref{an-1} using \eqref{poleforn-1} and the definition of shifted spinor-helicity variables:
\begin{align}
\langle \widehat{1}^In^J\rangle &=\frac{m}{\langle 2|\slashed{p}_1\cdot \slashed{p}_n|2\rangle}\left( \langle 2|p_1|n^J]\langle 21^I\rangle +\langle 2|p_n|1^I]\langle 2n^J\rangle +2m\langle 1^I2\rangle \langle 2n^J\rangle \right) \\
\langle 2|p_1|\widehat{2}]&=\langle 2|p_1|2] \,,  \qquad \langle 2|p_n|\widehat{2}]=\langle 2|p_n|2] +s_{1n} \,.
\end{align}
Finally we use the following identity
\begin{align}
 \left(\langle 2|p_1|2] +\langle 2|p_n|2] +s_{1n} \right)=s_{3,(n-1)}\,,
\end{align}
to derive the contribution of the last diagram $A^{(n-1)}$ as a function of the on-shell external variables:
\begin{align}
A^{(n-1)} =g^{(n-2)} \frac{\left( \langle 2|p_1|\textbf{n}]\langle 2\textbf{1}\rangle +\langle 2|p_n|\textbf{1}]\langle 2\textbf{n}\rangle +2m\langle \textbf{1}2\rangle \langle 2\textbf{n}\rangle  \right) ^2 }{s_{1n}\langle 23\rangle \langle 34\rangle \cdots \langle (n-2)(n-1)\rangle \left( \langle 2|\slashed{p}_1\cdot \slashed{p}_n|n-1\rangle +m^2\langle 2(n-1)\rangle \right)} \,. \label{n-1channel}
\end{align}
We combine the results of \eqref{archannel} and \eqref{n-1channel} to obtain a compact formula of the $n$-particle amplitude 
\begin{multline}
\mathcal{A}_n[\textbf{1},2^-,3^+,\ldots,\textbf{n}] =g^{n-2}\Bigg[ \tfrac{\left( \langle 2|p_1|\textbf{n}]\langle 2\textbf{1}\rangle +\langle 2|p_n|\textbf{1}]\langle 2\textbf{n}\rangle +2m\langle \textbf{1}2\rangle \langle 2\textbf{n}\rangle  \right) ^2 }{s_{1n}\langle 23\rangle \langle 34\rangle \cdots \langle (n-2)(n-1)\rangle \left( \langle 2|\slashed{p}_1\cdot \slashed{p}_n|n-1\rangle +m^2\langle 2(n-1)\rangle \right)} \\
+\sum _{r=3}^{n-2}\tfrac{\langle 2|\slashed{p}_{3,r}\cdot \prod _{k=r+1}^{n-2}\left\lbrace (s_{1\cdots k}-m^2)-\slashed{p}_k\cdot \slashed{p}_{1,k-1}\right\rbrace|n-1]\left( \langle 2|\slashed{p}_1.\slashed{p}_{3,r}|2\rangle \langle \textbf{1}\textbf{n}\rangle +p_{2,r}^2\langle \textbf{1}2\rangle \langle 2\textbf{n}\rangle \right)^2 \langle r(r+1)\rangle}{s_{23\ldots r}(s_{12\ldots r}-m^2)\ldots (s_{12\ldots (n-2)}-m^2) \langle 23\rangle \langle 34\rangle \ldots  \langle (n-2)(n-1)\rangle\langle 2|\slashed{p}_1.\slashed{p}_{2,r-1}|r\rangle\langle 2|\slashed{p}_1.\slashed{p}_{2,r}|r+1\rangle}\Bigg] \cr
\label{oneminus}  
\end{multline}
This is the main result of  this note.  While in principle, one could have attempted to  compute this $n$-point amplitude using BCFW recursion relations,  one would find that it requires use  of the recursion relations multiple times building from the known three-point on-shell amplitudes.  Instead we have shown that,  the massive-massless shift $[2^-\textbf{1}\rangle$ allows us to compute this amplitude using a single on-shell recursion that involves either the Parke-Taylor amplitudes or the massive vector boson amplitudes with all positive helicity gluons, derived in the previous section.  Our analysis demonstrates the potential of the covariant recursion relation introduced in \citep{Ballav:2020ese} to compute new classes of massive amplitudes.

As a simple check of the formula in equation \eqref{oneminus}, we have computed a few lower-point amplitudes by independent methods in appendix \ref{appendix-a.1} and showed that they agree with formula \eqref{oneminus}.  Additionally, in appendix \ref{appendix: b},  we treat \eqref{oneminus} as an ansatz and use the BCFW shift on the external gluon states to inductively prove the formula \eqref{oneminus}.

\subsection{Matching the NMHV amplitude in high energy limit}

We now consider the high energy limit of the scattering amplitude in \eqref{oneminus}.  This should reproduce the unique massless amplitudes for different helicity configurations since we used only the minimally coupled three-particle amplitudes \eqref{minimalthreepoint} as basic building blocks to construct the finite energy amplitude \citep{Arkani-Hamed:2017jhn}.  

The procedure of taking high energy limit of massive amplitudes is laid out in \citep{Arkani-Hamed:2017jhn} and further discussed in \citep{Ballav:2020ese}.  We do not repeat the procedure again but as a general rule of thumb, we show which component of massive spinor-helicity variables survives in this limit below
\begin{align}
|\textbf{n}\rangle \xrightarrow{p^0>>|\vec{p}|} |n^-\rangle\,, \qquad |\textbf{n}] \xrightarrow{p^0>>|\vec{p}|} |n^+] \quad \text{$\pm$ indicates helicity}\,.
\end{align}

The components of the finite energy amplitude \eqref{oneminus} in the high energy limit with  opposite helicity configurations for the pair of massive particles are non-vanishing due to the presence of both angle and square brackets of massive spinor-helicity variables and reproduce the correct MHV amplitudes as expected:
\begin{align}
\mathcal{A}_n^{\text{MHV}}[1^-,2^-,3^+,\ldots ,n^+]&=g^{n-2} \frac{\langle  12\rangle ^3}{\langle 23 \rangle \langle  34\rangle \cdots \langle  n1\rangle} \,,\\ 
\mathcal{A}_n^{\text{MHV}}[1^+,2^-,3^+,\ldots ,n^-]&=g^{n-2} \frac{\langle  2n\rangle ^4}{\langle 12\rangle \langle 23 \rangle \langle  34\rangle \cdots \langle  n1\rangle} \,.
\end{align}

The component of \eqref{oneminus} with positive helicity configuration for both massive particles vanishes explicitly as it should.  But the negative helicity configuration for both massive particles in the high energy limit should give us the NMHV amplitude. From \eqref{oneminus} we obtain the following result:
\eqst{
\mathcal{A}_n[1^-,2^-,3^+,\cdots, (n-1)^+,n^-]\cr =g^{n-2}\sum _{r=3}^{n-2}\tfrac{\langle 2|\slashed{p}_{3,r}\cdot \prod _{k=r+1}^{n-2}\left\lbrace s_{1\cdots k}-\slashed{p}_k\cdot \slashed{p}_{1,k-1}\right\rbrace \cdot \slashed{p}_{n-1}|n\rangle \left( \langle 2|\slashed{p}_1.\slashed{p}_{3,r}|2\rangle \langle 1n\rangle +p_{2,r}^2\langle 12\rangle \langle 2n\rangle \right)^2 \langle r(r+1)\rangle}{s_{23\ldots r}s_{12\ldots r}\ldots s_{12\ldots (n-2)}  \langle 23\rangle \langle 34\rangle \ldots \langle (n-1)n\rangle \langle 2|\slashed{p}_1.\slashed{p}_{2,r-1}|r\rangle\langle 2|\slashed{p}_1.\slashed{p}_{2,r}|r+1\rangle} \,. \label{NMHV1}
}
We simplify the product factor appearing in the numerator using the following identity: 
\begin{align}
\prod _{k=r+1}^{n-2}\left\lbrace s_{1\cdots k}-\slashed{p}_k\cdot \slashed{p}_{1,k-1}\right\rbrace \cdot \slashed{p}_{n-1}|n\rangle =\left(\prod _{k=r+1}^{n-2}s_{12\ldots k}\right)\left( p_{r+1}+\cdot +p_{n-1}\right)|n\rangle\,. \label{simpleid}
\end{align}
This identity can be derived from the one we have proved in Section \ref{masslessMHVchk}.  Furthermore, we use momentum conservation to get
\begin{align}
& \langle2|\slashed{p}_{3,r}.\slashed{p}_1|n\rangle+p_{2,r}^2\langle n2\rangle = \langle 2|\slashed{p}_{3,r}.(\slashed{p}_{r+1}+\ldots+\slashed{p}_{n-1})|n\rangle \,.
\end{align}
Substituting the above simplifications in \eqref{NMHV1},  we obtain:
\eqst{
\mathcal{A}_n[1^-,2^-,3^+,\cdots, (n-1)^+,n^-]\cr=g^{n-2}\sum _{r=3}^{n-2}\frac{\langle r(r+1)\rangle \langle 2|\slashed{p}_{3,r}\cdot \left(\slashed{p}_{r+1}+\cdots +\slashed{p}_{n-1}\right) |n\rangle ^3}{s_{23\ldots r}s_{12\ldots r} \langle 23\rangle \langle 34\rangle \ldots  \langle (n-1)n\rangle[1|p_{2,r-1}|r\rangle[1|p_{2,r}|r+1\rangle} \,. \label{NMHV}
}

We have obtained in \eqref{NMHV} a compact expression for the $n$-point NMHV amplitude that at first glance appears to be different from the standard expression in \cite{Dixon:2010ik}. Note that, the first term of the expression in equation \eqref{oneminus} (that corresponds to the last diagram in Figure \ref{flippedhelicitydiag}) does not contribute to the high energy limit since this involves massive spinor-helicity variables that do not survive in this limit.  It can be argued that this is a consequence of the massive-massless shift $[2^- \textbf{1}\rangle $ which we have used to derive this amplitude.  In a purely massless setup, one could use the BCFW shift $[1^- 2^- \rangle$, in which case the  last diagram in Figure \ref{flippedhelicitydiag} would certainly contribute. Therefore, in this case, the covariant massive-massless shift leads to a novel representation of the $n$-point NMHV amplitude.  In what follows, we will first take the soft limit of this amplitude to show that it obeys the Weinberg's soft theorem at leading order and subsequently we prove that the NMHV amplitude \eqref{NMHV} matches with the expression in \cite{Dixon:2010ik} for this specific ordering of external particles.

\subsubsection{Soft expansion of NMHV amplitude}

We begin with the NMHV expression in \eqref{NMHV} and take the limit $p_n \rightarrow 0$ of the gluon with momentum $p_n$.  In order to take the limit $p_n \rightarrow 0$,  we first scale the spinor-helicity variables as follows
\begin{align}
\lambda _{n\alpha} \longrightarrow \sqrt{\epsilon}  \lambda _{n\alpha} \,, \qquad \tilde{\lambda}_{n\dot{\alpha}} \longrightarrow \sqrt{\epsilon} \tilde{\lambda}_{n\dot{\alpha}} \,,
\end{align}
with $\epsilon \rightarrow 0$.  With this scaling, we find that the$r=n-2$ channel  of the NMHV amplitude in \eqref{NMHV} has the leading order contribution  $(\mathcal{O}\left(\frac{1}{\epsilon}\right))$ and the amplitude factorizes as follows 
  
\begin{align}
\lim _{p_n \rightarrow 0} \mathcal{A}_{n} &= \frac{[(n-1)1]}{[(n-1)n][n1]} \times \frac{\langle 12\rangle^3 }{\langle 23\rangle \langle 34\rangle \ldots \langle (n-1)1\rangle } \\
&=\frac{[(n-1)1]}{[(n-1)n][n1]} \times \mathcal{A}^{\text{MHV}}_{n-1}\,.
\end{align}
This follows from Weinberg soft theorem as well, as we will see now.

Using Weinberg soft theorem, we find that in the soft limit of the $n$-th gluon momentum, the  $n$-particle NMHV amplitude factorizes as a soft factor times an $(n-1)$-particle MHV amplitude as follows:
\begin{align}
\lim _{p_n \rightarrow 0}{\mathcal{A}}_n\left[1^-,2^-,3^+,.., (n-1)^+,n^-\right] =S^{(0)}(n^{-},(n-1)^+,1^-)\mathcal{A}_{n-1}\left[1^-,2^-,3^+,.., (n-1)^+\right]\,,
\end{align}
where the soft factor at leading order is given by \cite{PhysRev.140.B516,  Casali:2014xpa,  Strominger:2013lka}
\begin{align}
S^{(0)}(n^{-},(n-1)^+,1^-)=\left(\frac{\varepsilon _n^- \cdot p_{n-1}}{p_n\cdot p_{n-1}} -\frac{\varepsilon _n^- \cdot p_{1}}{p_n\cdot p_{1}} \right)\,.
\end{align}
Expressing the massless polarization vector in the spinor-helicity formalism
\begin{align}
\varepsilon_n^{-\mu}:=\frac{\langle n|\sigma ^\mu|q]}{[nq]} \,, 
\end{align}
and chosing the reference spinor as $|q]=|1]$,  we get the soft factor as follows
\begin{align}
S^{(0)}(n^{-},(n-1)^+,1^-)=\frac{[(n-1)1]}{[(n-1)n][n1]} \,.
\end{align}
So we have 
\begin{align}
{\mathcal{A}}_n\left[1^-,2^-,3^+,\cdots, (n-1)^+,n^-\right] =\frac{[(n-1)1]}{[(n-1)n][n1]} \mathcal{A}_{n-1}\left[1^-,2^-,3^+,\cdots, (n-1)^+\right]\,.  \label{softexpansion}
\end{align}
This is exactly what we get by the soft expansion of the NMHV amplitude in  \eqref{NMHV}.

\subsubsection{Matching the NMHV amplitude} 

Now that the preliminary check of the soft limit has been verified, we now show that the result in  \eqref{NMHV} matches exactly with the NMHV amplitude computed by Dixon et al. in \cite{Dixon:2010ik} for the specific ordering of negative helicity gluons that we have considered. The result in \cite{Dixon:2010ik} is of course more general in the sense that the positions of the two negative helicity gluons are completely arbitrary.  

In order to compare with our result \eqref{NMHV},  we begin with the result in \cite{Dixon:2010ik} and fix the positions of the two negative helicity particles as $1^{-}$, $2^{-}$.  The position of the third negative helicity particle is fixed to be $n^{-}$ in both of the results.  So the $n$-particle amplitude $\mathcal{A}_n[1^-,2^-,3^+,\ldots,(n-1)^+,n^-]$ (abbreviated as $\mathcal{A}_n^{\text{NMHV}}[1^-,2^-,n^-]$) from \cite{Dixon:2010ik} is given by
\begin{align}
\mathcal{A}_n^{\text{NMHV}}[1^-,2^-,n^-]&=\frac{1}{\langle12\rangle\langle 23\rangle\ldots\langle n1\rangle}\sum_{t=4}^{n-1} \mathcal{R}[n;2;t](\langle n1\rangle\langle n\,t\,2\,|\,1\rangle)^4\,.
\end{align}
The objects $\mathcal{R}[n;s;t]$ are defined to be 
\begin{align}
\mathcal{R}[n;s;t]:=\frac{1}{x_{st}^2}\frac{\langle s(s-1)\rangle}{\langle n\,t\,s\,|\,s\rangle\langle n\,t\,s\,|\,s-1\rangle}\frac{\langle t(t-1)\rangle}{\langle n\,s\,t\,|\,t\rangle\langle n\,s\,t\,|\,t-1\rangle}
\end{align}
with $\mathcal{R}[n;s;t]:=0$ for $t=s+1$ or $s=t+1$.
The spinor products are defined as 
\begin{align}
\langle n\,t\,s\,|\,s\rangle:=\langle n\,|x_{nt}\,x_{ts}\,|s\rangle
\end{align}
where 
\begin{align}
x_{st}^{\alpha\dot{\alpha}}:=(p_s+p_{s+1}+\ldots+p_{t-1})^{\alpha\dot{\alpha}}
\end{align}
for $s<t$, $x_{ss}=0$ and $x_{st}=-x_{ts}$ for $s>t$.
So we have 
\begin{align}
\mathcal{R}[n;2;t]:=\frac{1}{x_{2t}^2}\frac{\langle 21\rangle}{\langle n\,t\,2\,|\,2\rangle\langle n\,t\,2\,|\,1\rangle}\frac{\langle t(t-1)\rangle}{\langle n\,2\,t\,|\,t\rangle\langle n\,2\,t\,|\,t-1\rangle}
\end{align}
with $x_{2t}^2=(p_2+p_3+\ldots+p_{t-1})^2=s_{2(t-1)}$.
So the $n$-point NMHV gluon amplitude can be written as 

\begin{align}
\mathcal{A}_n^{\text{NMHV}}[1^-,2^-,n^-]&=\frac{\langle n1\rangle^3}{\langle 23\rangle\ldots\langle n1\rangle}\sum_{t=4}^{n-1}\frac{1}{s_{2(t-1)}}\frac{\langle t(t-1)\rangle\langle n\,t\,2\,|\,2\rangle^3}{\langle n\,t\,2\,|\,1\rangle\langle n\,2\,t\,|\,t\rangle\langle n\,2\,t\,|\,t-1\rangle}
\end{align}
Now by making a variable change, $t=r+1$ we can write 
\begin{align}
\mathcal{A}_n^{\text{NMHV}}[1^-,2^-,n^-]=&\sum_{r=3}^{n-2}\frac{\langle n1\rangle^3}{s_{23\ldots r}\langle 23\rangle\ldots\langle (r-1)r\rangle\langle (r+1)(r+2)\rangle\ldots\langle (n-1)n\rangle}\cr
&\hspace{1cm}\times\frac{\langle n\,(r+1)\,2\,|\,2\rangle^3}{\langle n\,(r+1)\,2\,|\,1\rangle\langle n\,2\,(r+1)\,|\,(r+1)\rangle\langle n\,2\,(r+1)\,|\,r\rangle}~.
\end{align}
The spinor products can be evaluated as follows 
\begin{align}
\langle n\,(r+1)\,2\,|\,1\rangle &=\langle n\,|x_{n(r+1)}\,x_{(r+1)2}\,|1\rangle\cr
&=\langle n\,|x_{(r+1)n}\,x_{2(r+1)}\,|1\rangle\cr
&=\langle n\,|(\slashed{p}_{r+1}+\slashed{p}_{r+2}+\ldots+\slashed{p}_{n-1}).\,(\slashed{p}_2+\slashed{p}_3+\ldots+\slashed{p}_r)\,|1\rangle\cr
&=\langle n\,|(\slashed{p}_{r+1}+\slashed{p}_{r+2}+\ldots+\slashed{p}_{n-1}+\slashed{p}_n).\,(\slashed{p}_1+\slashed{p}_2+\ldots+\slashed{p}_r)\,|1\rangle\cr
&=-\langle n\,|(p_1+p_2+\ldots+p_r)^2\,|1\rangle\cr
&=-s_{12\ldots r}\langle n1\rangle
\end{align}
Similarly we simplify other spinor products and obtain
\begin{align}
& \langle n\,2\,(r+1)\,|\,r\rangle=\langle n1 \rangle [1|p_{2,(r-1)}|r\rangle\,,\\
& \langle n\,2\,(r+1)\,|\,(r+1)\rangle=\langle n1 \rangle [1|p_{2,r}|(r+1)\rangle\,,\\
& \langle n\,(r+1)\,2\,|\,2\rangle=\langle 2|\slashed{p}_{3,r}.(\slashed{p}_{r+1}+\ldots+\slashed{p}_{n-1})|n\rangle=\langle2|\slashed{p}_{3,r}.\slashed{p}_1|n\rangle+p_{2,r}^2\langle n2\rangle.
\end{align}
Assembling all these, we can rewrite the NMHV amplitude as follows 
\begin{align}
\mathcal{A}_n^{\text{NMHV}}[1^-,2^-,n^-]=\sum _{r=3}^{n-2}\tfrac{\langle 2|\slashed{p}_{3,r}\cdot \left( \slashed{p}_{r+1}+\cdots +\slashed{p}_{n-1}\right) |n\rangle^3 }{s_{23\ldots r}s_{12\ldots r} \langle 23\rangle \langle 34\rangle \ldots \langle (r-1)\,r\rangle \langle (r+1)(r+2)\rangle \ldots  \langle (n-1)n\rangle[1|p_{2,r-1}|r\rangle[1|p_{2,r}|r+1\rangle} \,. 
\end{align}
This exactly matches with the NMHV amplitude in equation \eqref{NMHV}. We conclude that the massive amplitude we computed has the expected high energy limit.

\subsection{Spurious poles}
Although the covariant recursion allows us to determine the $n$-particle amplitude in a compact form, in the case of $n\, \geq\, 6$ the final expression in \eqref{oneminus} contains spurious poles which are not associated to any propagator going on-shell.  These poles are arising in the form of $\langle 2|\slashed{p}_1\cdot \slashed{p}_{2,r-1}|r\rangle$ and $\langle 2|\slashed{p}_1\cdot \slashed{p}_{2,r}|r+1\rangle$ in the denominator of the expression in \eqref{oneminus}.
 Any on-shell recursion scheme will generically be infected with such spurious poles as the manifest locality is sacrificed at the altar of staying on-shell.  In the case of BCFW recursion relations for massless theories such as non-Abelian gauge theory, the spurious poles have been analysed extensively. These poles are not physical and their final contribution to the amplitude (via residue) vanishes \cite{Hodges:2009hk}. 
 
 We expect that the same should be true in the present case as the theories under considerations are local. However, as is well known,  proving that spurious poles are indeed spurious is no easy task even for scattering amplitudes of massless particles and the proofs usually involve additional tools such as momentum twistor variables \cite{Hodges:2009hk}.  We do not pursue this important question in the present work but give a small evidence that the poles which arise in \eqref{oneminus} and that do not correspond to on-shell propagators are indeed spurious. 
 
 We consider the six-point amplitude and  evaluate it using $[\textbf{6}5^+\rangle$ shift, which leads to the following expression for the six-point amplitude: 
\begin{align}
\mathcal{A}_6[\textbf{1},2^- &,3^+ ,4^+,5^+,\textbf{6}] = g^4\Bigg[\tfrac{\left( \langle \textbf{6}|p_2|3]\langle 2\textbf{1}\rangle + \langle 2|p_1|3]\langle \textbf{6}\textbf{1}\rangle \right)^2 [34]\langle 4|p_6|5]}{[23]\langle 54\rangle(s_{12}-m^2)(s_{123}-m^2)(s_{56}-m^2)\left( \langle 2|\slashed{p}_1\cdot \slashed{p}_6|4\rangle +m^2 \langle 24\rangle \right)} \cr
&+ \tfrac{\left( \langle 4|p_6|5]\left\lbrace \langle 2\textbf{1}\rangle  \langle 2|p_1|\textbf{6}]+\langle 2|p_6|\textbf{1}]\langle 2\textbf{6}\rangle  +2m \langle 2\textbf{1}\rangle \langle \textbf{6}2\rangle \right\rbrace+\langle 2\textbf{1}\rangle [\textbf{6}5]\langle 4|\slashed{p}_5\cdot \slashed{p}_1|2\rangle   +\langle 4|p_5|\textbf{1}]\langle 2|p_6|5]  \langle 2\textbf{6}\rangle   \right) ^2 }{\langle 23\rangle \langle 34\rangle \langle 45\rangle \left( \langle 4|\slashed{p}_5\cdot \slashed{p}_1\cdot \slashed{p}_6|5]+ s_{16}\langle 4|p_6|5]\right) (s_{56}-m^2)\left( \langle 2|\slashed{p}_1\cdot (\slashed{p}_5+\slashed{p}_6)|4\rangle +m^2 \langle 24\rangle \right)}  \cr
&+\tfrac{\left( \langle 2|p_1+p_6|2][5|\slashed{p}_1\cdot \slashed{p}_6|5]-s_{16}[5|\slashed{p}_2\cdot \slashed{p}_6|5]\right) \langle \textbf{1}\textbf{6}\rangle ^2 \langle 2|p_1+p_6|5] ^2 }{s_{156}s_{16}\langle 23\rangle \langle 34\rangle \left( [5|\slashed{p}_6\cdot \slashed{p}_1|2]+m^2[52]\right) \left( \langle 4|\slashed{p}_5\cdot \slashed{p}_1 \cdot \slashed{p}_6|5] +s_{16}\langle 4|p_6|5]\right) } \Bigg]\,.
\end{align}
Here the spurious pole condition is the following
\begin{align}
\langle 4|\slashed{p}_5\cdot \slashed{p}_1 \cdot \slashed{p}_6|5] +s_{16}\langle 4|p_6|5]=0\,.
\end{align}
From our original shift $[2^-\textbf{1}\rangle$,  the $6$-point amplitude can be written using \eqref{oneminus} in the following form 
\begin{align}
\mathcal{A}_6[\textbf{1},2^-,3^+,4^+,5^+,\textbf{6}] =&\, g^4\Bigg[\tfrac{\left( \langle 2|\slashed{p}_1.\slashed{p}_{3}|2\rangle \langle \textbf{1}\textbf{6}\rangle +p_{2,3}^2\langle \textbf{1}2\rangle \langle 2\textbf{6}\rangle \right)^2 \langle 2|\slashed{p}_{3}\cdot \left((s_{56}-m^2)-\slashed{p}_4\cdot \slashed{p}_{1,3} \right)|5]}{s_{23}(s_{123}-m^2)(s_{56}-m^2)\langle 23\rangle \langle 45\rangle  \langle 2|\slashed{p}_1.\slashed{p}_{2}|3\rangle \langle 2|\slashed{p}_1.\slashed{p}_{2,3}|4\rangle}  \cr
&\hspace{6mm}+\tfrac{\left( \langle 2|\slashed{p}_1.\slashed{p}_{3,4}|2\rangle \langle \textbf{1}\textbf{6}\rangle +p_{2,4}^2\langle \textbf{1}2\rangle \langle 2\textbf{6}\rangle \right)^2 \langle 2|p_{3,4}|5]}{s_{234}(s_{56}-m^2)\langle 23\rangle \langle 34\rangle  \langle 2|\slashed{p}_1.\slashed{p}_{2,3}|4\rangle\langle 2|\slashed{p}_1.\slashed{p}_{2,4}|5\rangle}  \cr
&\hspace{6mm}+\tfrac{\left( \langle 2\textbf{1}\rangle \langle 2|p_1|\textbf{6}] +\langle 2|p_6|\textbf{1}]\langle 2\textbf{6}\rangle +2m \langle 2\textbf{1}\rangle \langle \textbf{6}2\rangle \right) ^2 }{s_{16}\langle 23\rangle \langle 34\rangle \langle 45\rangle \left( \langle 2|\slashed{p}_1\cdot \slashed{p}_6|5\rangle +m^2 \langle 25\rangle \right)}\Bigg]\,.
\end{align}
The spurious pole in the above amplitude is given by the following condition
\begin{align}
\langle 2|\slashed{p}_1.(\slashed{p}_{2}+\slashed{p}_3)|4\rangle=0 \,.
\end{align}
 It is easy to check that both expressions for the six-point amplitude contain the same set of physical poles. However we see that these have different spurious poles. In particular, when the spurious pole condition is satisfied for one expression, the other one is finite. As both are representations of the same scattering amplitude, we conclude that at least in this simple example, the residues of the spurious poles indeed sum up to zero.


\section{Summary and Outlook}\label{summary}
In this note, we applied the covariant recursion relations introduced in \cite{Ballav:2020ese} to compute scattering amplitudes with massive particles which hitherto were not known in the literature. The class of amplitudes we chose to focus on are massive analogues of the MHV and NMHV amplitudes in Yang-Mills theory.  In the high energy limit, these two classes indeed reduce to the MHV and NMHV amplitudes respectively. Our work can thus be considered as mirroring the computation of MHV and NMHV amplitudes in gauge theory using BCFW recursion relation. 

The analogue of the NMHV amplitude  consists of two massive vector bosons, one negative helicity gluon (that is colour adjacent to the massive bosons) and remaining positive helicity gluons. We showed that for this class of amplitudes, the massive-massless shift leads to a remarkably simple computation and we could generate a compact,  little group covariant formula for the final amplitude by using a single recursion.  Interestingly we have shown that given the final form for the amplitude derived using the covariant recursion techniques,  one can verify that our result indeed satisfies the BCFW recursion relation. This is shown in detail in Appendix \ref{appendix: b}.  

It is useful to recall the two key ingredients that went into this computation. First of all we derive the scattering amplitude involving a pair of massive vector bosons and only positive helicity gluons by relating it to an amplitude involving two massive scalars and positive helicity gluons. While this relation was derived previously in \cite{Boels:2011zz},  we work with  little group covariant expressions and provide an inductive proof of the scattering amplitude by making use of the covariant recursion (in the massive spinor-helicity formalism).  This was then used as an input to calculate the scattering amplitude in which we flip the helicity of the gluon adjacent to one of the massive particles. We use a specific massive-massless shift such that the resulting subamplitudes in the covariant recursion involved either pure gluon amplitudes or amplitudes involving massive vector bosons and positive helicity gluons. This led to a compact expression for the relevant scattering amplitude in \eqref{oneminus}, which is the main result of this work. 

We checked  the correctness of our result  by taking the high energy limit and showing that they reduce to the expected MHV and the NMHV amplitudes. As mentioned previously we also checked that the $n$-point amplitude satisfies the usual BCFW recursion relation. Interestingly, our representation of the NMHV amplitude obtained in the high energy limit is not identical to the one obtained previously in  \citep{Dixon:2010ik}.  We showed that the two expressions are equal and it will be interesting to study the representation for NMHV amplitude that we obtained in more detail in its own right.

In this note, we have restricted ourselves to a particular configuration in which the position of the negative helicity gluon is adjacent to the massive vector bosons. But in fact it is possible to make the position of the negative helicity gluon completely arbitrary and use the covariant massive-massless shift or the BCFW shift in combination with the amplitudes calculated in this work to derive these  scattering amplitudes. One could also include additional negative helicity gluons and systematically proceed to calculate the resulting scattering amplitudes. 
However in order to compute amplitudes with more than two massive particles using the covariant recursion relations, one would require knowledge of a wider class of amplitudes. 
We hope to address these issues in the future.

\vspace{.5cm}
\section*{Acknowledgements}

We are grateful to Alok Laddha for suggesting the problem, numerous illuminating discussions and valuable comments on the draft. We thank Sujay Ashok for constant support, discussions and valuable feedback on the draft. SB thanks Arnab Rudra and NCTS Taiwan for the support. SB is supported by the grant INST/PHY/2020019.

\begin{appendix}

\section{Lower-point amplitudes} \label{appendix:A}
In this section, we show that the four- and five-point amplitudes involving massive vector bosons computed previously in \cite{Ballav:2020ese} using the recursion relations with covariant massive-massless shift,  are consistent with the general formula \eqref{nptvector} and \eqref{oneminus} derived in this work.  
\subsection{Lower-point amplitudes with identical gluons}
\label{appendix: a}
As mentioned at the beginning of section \ref{section-3.1},  the relevant amplitudes needed to set up the method of induction are given as follows \citep{Arkani-Hamed:2017jhn,  Ballav:2020ese}
\begin{align}
\mathcal{A}_4\left[\textbf{1},2^+,3^+,\textbf{4}\right] &=g^2\frac{[23] \langle \textbf{1}\textbf{4}\rangle ^2}{\langle 23\rangle(s_{12}-m^2)}\,.\label{fourptvector} \\
\mathcal{A}_5\left[\textbf{1},2^+,3^+,4^+,  \textbf{5}\right]
&=g^3\frac{\langle \textbf{1}\textbf{5}\rangle ^2[2|\slashed{p}_1 \cdot(\slashed{p}_2+\slashed{p}_3)|4]}{\langle 23\rangle \langle 34\rangle  (s_{12}-m^2) (s_{45}-m^2)}\,.  \label{fiveptvector}
\end{align}
Although, the four-particle amplitude matches straightforwardly with the expression that we obtain from the general formula \eqref{nptvector} with $n=4$,  but the five-particle amplitude \eqref{fiveptvector} does not identically match with the expression that we get from \eqref{nptvector}.  In order to match these two expressions, we now prove the following identity:
\begin{align}
[2|\slashed{p}_1 \cdot (\slashed{p}_{2}+\slashed{p}_3)|4]=[2|\left\lbrace(s_{123}-m^2)-\slashed{p}_3\cdot (\slashed{p}_1+\slashed{p}_2)\right\rbrace |4] \,. \label{fiveptid}
\end{align}
We rewrite the R.H.S in the following way
\begin{align}
[2|\left\lbrace(s_{123}-m^2)-\slashed{p}_3\cdot (\slashed{p}_1+\slashed{p}_2)\right\rbrace |4]=2p_3\cdot p_{1,2}[24]-[2|\slashed{p}_3\cdot \slashed{p}_{1,2}|4]+2p_1\cdot p_2[24]\,. \label{t1}
\end{align}
Using the following identity satisfied by the Pauli matrices (and identity matrix) 
\begin{align}
\Big(\sigma ^\mu \bar{\sigma} ^\nu +\sigma ^\nu \bar{\sigma} ^\mu \Big)_{\dot{\alpha}}~^{\dot{\beta}}=2 \eta ^{\mu \nu} \delta _{\dot{\alpha}}~^{\dot{\beta}}\,,  \qquad \text{where}\quad \bar{\sigma}_\mu ^{\dot{\alpha}\alpha}=\epsilon ^{\dot{\alpha}\dot{\beta}} \epsilon ^{\alpha \beta} \sigma _{\mu \beta \dot{\beta}}\,; \label{Pauliid}
\end{align}
we get
\begin{align}
2p_3\cdot p_{1,2}[24]=[2|\slashed{p}_3\cdot \slashed{p}_{1,2}|4]+[2| \slashed{p}_{1,2}\cdot \slashed{p}_3|4] \,, \qquad 2p_1\cdot p_2[24]=[2|\slashed{p}_1\cdot \slashed{p}_2|4]\,.
\end{align}
In the last equality, we use $p_2|2]=0$.  Substituting these results in \eqref{t1}, we easily obtain the identity \eqref{fiveptid}.
This completes the check of the formula \eqref{nptvector} for $n$-particle amplitude involving a pair of massive vector bosons and positive helicity gluons for lower-point amplitudes.

\subsection{Lower-point amplitudes with helicity flip}\label{appendix-a.1}
In this section, we verify the formula for the $n$-particle amplitude \eqref{oneminus} involving a pair of massive vector bosons,  one minus helicity gluon which is colour adjacent to the massive particles and $(n-3)$ positive helicity gluons for $n=4 $ and $5$.  First we write down the four- and five-point amplitudes directly by using \eqref{oneminus} and then compare with the amplitudes computed using other techniques like unitarity and recursion involving massless-massless shift.

\subsubsection{Four-point amplitude}
Let us start with the four-particle amplitude for which only the first term in \eqref{oneminus} contributes
\begin{align}
\mathcal{A}_4\left[\textbf{1},2^-,3^+,\textbf{4}\right]=g^2 \frac{\left( \langle 2|p_1|\textbf{4}]\langle 2\textbf{1}\rangle +\langle 2|p_4|\textbf{1}]\langle 2\textbf{4}\rangle +2m\langle \textbf{1}2\rangle \langle 2\textbf{4}\rangle  \right) ^2 }{s_{14}\langle 23\rangle \left( \langle 2|\slashed{p}_1\cdot \slashed{p}_4|3\rangle +m^2\langle 23\rangle \right)} \,.
\end{align}
We simplify the following terms using momentum conservation
\begin{align}
& \langle 2|p_1|\textbf{4}]=-\langle 2|p_3|\textbf{4}]-m\langle 2\textbf{4}\rangle \,, \quad \langle 2|p_4|\textbf{1}]=-\langle 2|p_3|\textbf{1}]-m\langle 2\textbf{1}\rangle \,,\\
& \langle 2|\slashed{p}_1\cdot \slashed{p}_4|3\rangle +m^2\langle 23\rangle =-(s_{12}-m^2)\langle 23\rangle \,,
\end{align}
and express the four-particle amplitude in the following form
\begin{align}
\mathcal{A}_4\left[\textbf{1},2^-,3^+,\textbf{4}\right]=g^2 \frac{\left( [3\textbf{4}]\langle 2\textbf{1}\rangle +[3\textbf{1}]\langle 2\textbf{4}\rangle  \right) ^2 }{s_{23} \left( s_{12}-m^2 \right)} \,.
\end{align}
This result matches exactly with amplitudes computed in \citep{Arkani-Hamed:2017jhn,  Ballav:2020ese}.  Next we move to five-particle amplitude which we compute using massless-massless shift.  
\subsubsection{Five-point amplitude}
\begin{figure}
\includegraphics[scale=.38]{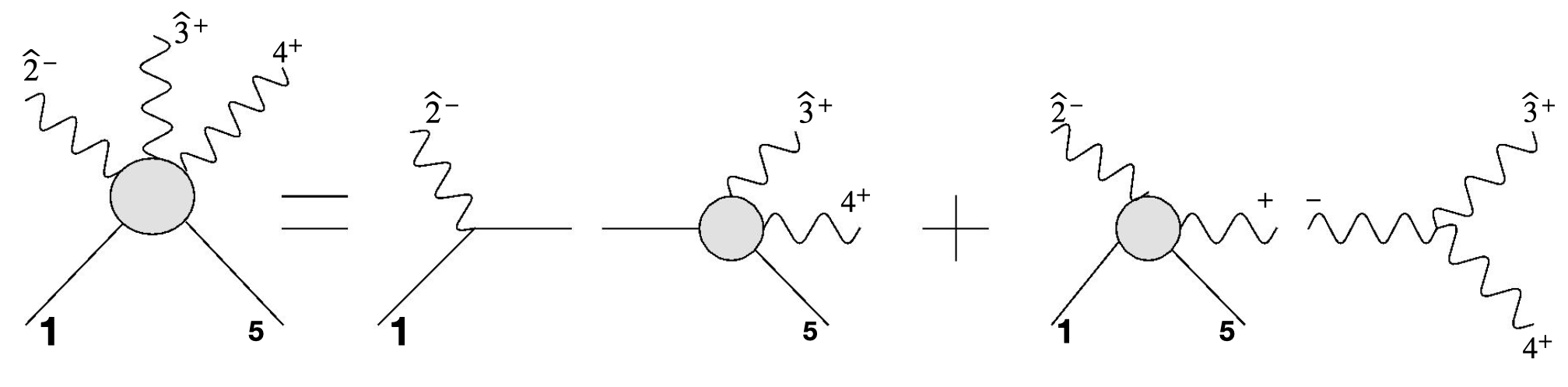}
\caption{Scattering channels to compute $\mathcal{A}_5\left[\textbf{1},2^-,3^+,4^+,\textbf{5}\right]$ with $[2^-3^+\rangle$ shift} 
\label{fiveptbcfw}
\end{figure}

We use the $[2^-3^+\rangle$ massless-massless shift to calculate the colour-ordered five-particle amplitude.  The scattering channels to evaluate this amplitude using the particular shift are given in Figure \ref{fiveptbcfw}.
We consider the following shift for massless spinor-helicity variables  
\begin{align}
[\widehat{2}|=[2|-z[3|\,, \qquad |\widehat{3}\rangle =|3\rangle +z|2\rangle \,. \label{masslessspinorshift}
\end{align}
The contribution to five-particle amplitude from the first diagram is obtained by gluing the four-particle amplitude alongwith the three-particle amplitude \eqref{minimalthreepoint} for negative helicity gluon and unshifted propagator $\frac{1}{s_{12}-m^2}$
\begin{align}
\mathcal{A}_5^I\left[\textbf{1},2^-,3^+,4^+,\textbf{5}\right]=\frac{g^3}{m^2}\frac{\langle 2|p_1|3]}{[23]}\frac{[34]\langle \textbf{5}|\widehat{I}|\textbf{1}]^2}{[23]\langle \widehat{3}4\rangle (s_{12}-m^2)(s_{45}-m^2)}\,.
\end{align}
We get the pole $z_{(12)}$ for first diagram by setting the shifted propagator $\frac{1}{\widehat{s}_{12}-m^2}$ on-shell
\begin{align}
z_{(12)}=\frac{\langle 2|p_1|2]}{\langle 2|p_1|3]}\,.
\end{align}
Using momentum conservation and definition of shifted massless spinor-helicity variables of \eqref{masslessspinorshift},  we get get rid of the dependence on internal momentum and evaluate remaining shifted spinor products at this pole.   
We obtain the contribution of first diagram
\begin{align}
\mathcal{A}_5^I\left[\textbf{1},2^-,3^+,4^+,\textbf{5}\right]&=g^3\frac{\left(\langle \textbf{5}|p_2|3]\langle 2\textbf{1}\rangle +\langle 2|p_1|3]\langle \textbf{5}\textbf{1}\rangle  \right)^2[34]}{[23](s_{12}-m^2)(s_{45}-m^2)\left(\langle 2|\slashed{p}_1\cdot \slashed{p}_5|4\rangle +m^2 \langle 24\rangle \right) } \cr
&=g^3\frac{\left( \langle 2|\slashed{p}_1\cdot \slashed{p}_3|2\rangle \langle \textbf{1}\textbf{5}\rangle +p_{23}^2\langle \textbf{1}2\rangle \langle 2\textbf{5}\rangle  \right)^2   \langle 2|p_3|4]}{ s_{23}(s_{123}-m^2)\langle 23\rangle \langle 2|\slashed{p}_1\cdot \slashed{p}_2|3\rangle \langle 2|\slashed{p}_1\cdot \slashed{p}_{2,3}|4\rangle }\,.  \label{fivept-firstdiag}
\end{align}
According to the formula \eqref{oneminus},  there exists two scattering channels contributing to the five-particle amplitude.  For $n=5$, the sum in the second term of \eqref{oneminus} becomes a single term which matches exactly with above expression.

The contribution from the second diagram in Figure \ref{fiveptbcfw} is obtained by gluing the two subamplitudes along with unshifted propagator $\frac{1}{s_{34}}$.  After evaluating the shifted spinor products at $z_{(34)}=\frac{\langle 34\rangle}{\langle 24\rangle}$, we get the contribution from this diagram as follows
\begin{align}
\mathcal{A}_5^{II}\left[\textbf{1},2^-,3^+,4^+,\textbf{5}\right]=g^3 \frac{\left(\langle 2\textbf{1}\rangle \langle 2|p_1|5]+\langle 2|p_5|\textbf{1}]\langle 2\textbf{5}\rangle +2m\langle 2\textbf{1}\rangle \langle \textbf{5}2\rangle \right)^2}{\langle 23\rangle \langle 34\rangle s_{15}\left( \langle 2|\slashed{p}_1\cdot \slashed{p}_5|4\rangle +m^2 \langle 24\rangle \right)}\,. \label{fivept-seconddiag}
\end{align}
This expression matches with the first term in \eqref{oneminus} with $n=5$.

\section{Flip helicity amplitude from BCFW recursion}
\label{appendix: b}
In this section, we present an inductive proof of the formula in \eqref{oneminus} using the BCFW recursion.
To set up the induction, we first of all ensure that the four- and five-point amplitudes that have been calculated previously  in Appendix \ref{appendix-a.1} using the BCFW recursion are consistent with the general expression.
\begin{figure}
\includegraphics[scale=.35]{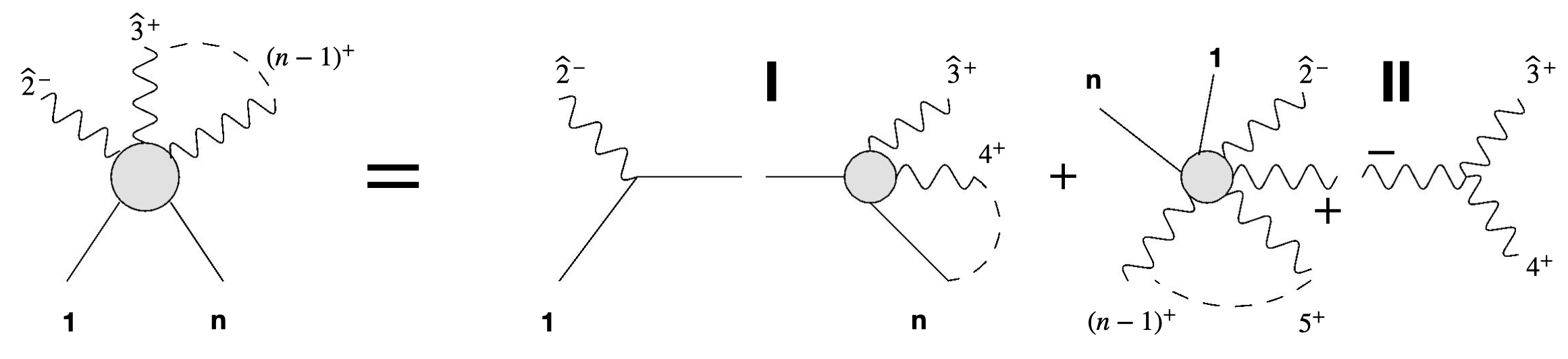}
\caption{Pictorial representation of BCFW recursion with $[2^-3^+\rangle$ shift. }
\label{masslesscheck}
\end{figure}

Given the match of the lower-point amplitudes we now assume that the expression \eqref{nptvector} is true for $(n-1)$-particle amplitude and use this to construct $n$-particle amplitude. We use the $[2^-3^+\rangle$ shift which corresponds to the shifts of the following spinor-helicity variables:
\begin{align}
|\widehat{2}]=|2]+z|3]\,, \qquad |\widehat{3}\rangle =|3\rangle -z|2\rangle \,. 
\end{align}
With this shift, the possible channels that have a non-zero contribution to the $n$-point amplitude are shown in Figure \ref{masslesscheck}.
The first diagram contributes to the $n$-point amplitude as 
\begin{align}
\mathcal{A}_n^I&=\mathcal{A}_L[\textbf{1},\widehat{2}^-,\widehat{\textbf{I}}] \frac{1}{s_{12}-m^2}\mathcal{A}_R[\widehat{\textbf{I}},\widehat{3}^+,\ldots,(n-1)^+, \textbf{n}]\,. 
\end{align}

Substituting the $3$-point amplitude as given in \eqref{minimalthreepoint} and the expression for the right subamplitude from \eqref{nptvector} and evaluating shifted spinor products at the simple pole 
\begin{align}
z_{(12)}=-\frac{\langle 2|p_1|2]}{\langle 2|p_1|3]}\,,
\end{align}
we obtain
\begin{align}
\label{A1}
\mathcal{A}_n^I&=\tfrac{\langle 2|\slashed{p}_{3}\cdot \prod _{k=4}^{n-2}\left\lbrace (s_{1\cdots k}-m^2)-\slashed{p}_k\cdot \slashed{p}_{1,k-1}\right\rbrace|n-1]\left( \langle 2|\slashed{p}_1.\slashed{p}_{3}|2\rangle \langle \textbf{1}\textbf{n}\rangle +p_{2,3}^2\langle \textbf{1}2\rangle \langle 2\textbf{n}\rangle \right)^2 \langle 34\rangle}{s_{23}(s_{123}-m^2)\ldots (s_{12\ldots (n-2)}-m^2) \langle 23\rangle \langle 34\rangle \ldots  \langle (n-2)(n-1)\rangle\langle 2|\slashed{p}_1.\slashed{p}_{2}|3\rangle\langle 2|\slashed{p}_1.\slashed{p}_{2,3}|4\rangle} \,.
\end{align}

The second diagram contributes to the $n$-point amplitude as 
 \begin{align}
\mathcal{A}_n^{II}&=\mathcal{A}_L[\textbf{1},\widehat{2}^-,\widehat{I}^+,5^+,\ldots,(n-1)^+,\textbf{n}] \frac{1}{s_{34}}\mathcal{A}_R[\widehat{I}^-,\widehat{3}^+,4^+] 
\end{align}

We substitute the left subamplitude from the expression in \eqref{nptvector} by assuming that it is true for $(n-1)$-point amplitude. The right subamplitude is a pure gluon amplitude and is given by the Parke-Taylor formula.  Using these expressions and  simplifying further we get
\begin{multline}
\mathcal{A}_n^{II}=g^{n-2}\Bigg[ \tfrac{\left( \langle 2|p_1|\textbf{n}]\langle 2\textbf{1}\rangle +\langle 2|p_n|\textbf{1}]\langle 2\textbf{n}\rangle +2m\langle \textbf{1}2\rangle \langle 2\textbf{n}\rangle  \right) ^2 }{s_{1n}\langle 23\rangle \langle 34\rangle \cdots \langle (n-2)(n-1)\rangle \left( \langle 2|\slashed{p}_1\cdot \slashed{p}_n|n-1\rangle +m^2\langle 2(n-1)\rangle \right)} \\
+\sum _{r=4}^{n-2}\tfrac{\langle 2|\slashed{p}_{3,r}\cdot \prod _{k=r+1}^{n-2}\left\lbrace (s_{1\cdots k}-m^2)-\slashed{p}_k\cdot \slashed{p}_{1,k-1}\right\rbrace|n-1]\left( \langle 2|\slashed{p}_1.\slashed{p}_{3,r}|2\rangle \langle \textbf{1}\textbf{n}\rangle +p_{2,r}^2\langle \textbf{1}2\rangle \langle 2\textbf{n}\rangle \right)^2 \langle r(r+1)\rangle}{s_{23\ldots r}(s_{12\ldots r}-m^2)\ldots (s_{12\ldots (n-2)}-m^2) \langle 23\rangle \langle 34\rangle \ldots  \langle (n-2)(n-1)\rangle\langle 2|\slashed{p}_1.\slashed{p}_{2,r-1}|r\rangle\langle 2|\slashed{p}_1.\slashed{p}_{2,r}|r+1\rangle}\Bigg]\,.
\label{A2}
\end{multline}

Combining the contributions from two diagrams \eqref{A1} and \eqref{A2}, we get the $n$-particle amplitude $\mathcal{A}_n[\textbf{1},2^-,3^+,\ldots,\textbf{n}] $ which exactly matches with \eqref{oneminus}. This completes alternative check of \eqref{oneminus} using BCFW recursion.

\end{appendix}

\bibliographystyle{hephys}
\bibliography{Self}

\begin{thebibliography}{10}
\newcommand{\enquote}[1]{``#1''}

\bibitem{PhysRevLett.56.2459}
S.~J. Parke and T.~R. Taylor, \enquote{Amplitude for $n$-Gluon Scattering},
  \href{http://dx.doi.org/10.1103/PhysRevLett.56.2459}{\emph{Phys. Rev. Lett.}
  \textbf{56} (1986) 2459}.

\bibitem{GUNION1985333}
J.~Gunion and Z.~Kunszt, \enquote{Improved analytic techniques for tree graph
  calculations and the ggqqℓℓ subprocess},
  \href{http://dx.doi.org/https://doi.org/10.1016/0370-2693(85)90774-9}{\emph{Physics
  Letters B} \textbf{161[4]} (1985) 333},
  \href{https://www.sciencedirect.com/science/article/pii/0370269385907749}{{\tt
  URL}}.

\bibitem{Xu:1986xb}
Z.~Xu, D.-H. Zhang and L.~Chang, \enquote{{Helicity Amplitudes for Multiple
  Bremsstrahlung in Massless Nonabelian Gauge Theories}},
  \href{http://dx.doi.org/10.1016/0550-3213(87)90479-2}{\emph{Nucl. Phys. B}
  \textbf{291} (1987) 392}.

\bibitem{Gastmans:1990xh}
R.~Gastmans and T.~T. Wu, {The Ubiquitous photon: Helicity method for QED and
  QCD}, volume~80 1990.

\bibitem{Dixon:1996wi}
L.~J. Dixon, \enquote{{Calculating scattering amplitudes efficiently}},
  \emph{in} {Theoretical Advanced Study Institute in Elementary Particle
  Physics (TASI 95): QCD and Beyond}, p. 539--584 1996,
  \href{http://arxiv.org/abs/hep-ph/9601359}{{\tt arXiv:hep-ph/9601359}}.

\bibitem{Britto:2004ap}
R.~Britto, F.~Cachazo and B.~Feng, \enquote{{New recursion relations for tree
  amplitudes of gluons}},
  \href{http://dx.doi.org/10.1016/j.nuclphysb.2005.02.030}{\emph{Nucl. Phys. B}
  \textbf{715} (2005) 499}, \href{http://arxiv.org/abs/hep-th/0412308}{{\tt
  arXiv:hep-th/0412308}}.

\bibitem{Britto:2005fq}
R.~Britto, F.~Cachazo, B.~Feng and E.~Witten, \enquote{{Direct proof of
  tree-level recursion relation in Yang-Mills theory}},
  \href{http://dx.doi.org/10.1103/PhysRevLett.94.181602}{\emph{Phys. Rev.
  Lett.} \textbf{94} (2005) 181602},
  \href{http://arxiv.org/abs/hep-th/0501052}{{\tt arXiv:hep-th/0501052}}.

\bibitem{Schwinn:2005pi}
C.~Schwinn and S.~Weinzierl, \enquote{{Scalar diagrammatic rules for Born
  amplitudes in QCD}},
  \href{http://dx.doi.org/10.1088/1126-6708/2005/05/006}{\emph{JHEP}
  \textbf{05} (2005) 006}, \href{http://arxiv.org/abs/hep-th/0503015}{{\tt
  arXiv:hep-th/0503015}}.

\bibitem{Franken:2019wqr}
R.~Franken and C.~Schwinn, \enquote{{On-shell constructibility of Born
  amplitudes in spontaneously broken gauge theories}},
  \href{http://dx.doi.org/10.1007/JHEP02(2020)073}{\emph{JHEP} \textbf{02}
  (2020) 073}, \href{http://arxiv.org/abs/1910.13407}{{\tt arXiv:1910.13407
  [hep-th]}}.

\bibitem{Badger:2005zh}
S.~Badger, E.~Glover, V.~Khoze and P.~Svrcek, \enquote{{Recursion relations for
  gauge theory amplitudes with massive particles}},
  \href{http://dx.doi.org/10.1088/1126-6708/2005/07/025}{\emph{JHEP}
  \textbf{07} (2005) 025}, \href{http://arxiv.org/abs/hep-th/0504159}{{\tt
  arXiv:hep-th/0504159}}.

\bibitem{Badger:2005jv}
S.~Badger, E.~Glover and V.~V. Khoze, \enquote{{Recursion relations for gauge
  theory amplitudes with massive vector bosons and fermions}},
  \href{http://dx.doi.org/10.1088/1126-6708/2006/01/066}{\emph{JHEP}
  \textbf{01} (2006) 066}, \href{http://arxiv.org/abs/hep-th/0507161}{{\tt
  arXiv:hep-th/0507161}}.

\bibitem{Ferrario:2006np}
P.~Ferrario, G.~Rodrigo and P.~Talavera, \enquote{{Compact multigluonic
  scattering amplitudes with heavy scalars and fermions}},
  \href{http://dx.doi.org/10.1103/PhysRevLett.96.182001}{\emph{Phys. Rev.
  Lett.} \textbf{96} (2006) 182001},
  \href{http://arxiv.org/abs/hep-th/0602043}{{\tt arXiv:hep-th/0602043}}.

\bibitem{Schwinn:2007ee}
C.~Schwinn and S.~Weinzierl, \enquote{{On-shell recursion relations for all
  Born QCD amplitudes}},
  \href{http://dx.doi.org/10.1088/1126-6708/2007/04/072}{\emph{JHEP}
  \textbf{04} (2007) 072}, \href{http://arxiv.org/abs/hep-ph/0703021}{{\tt
  arXiv:hep-ph/0703021}}.

\bibitem{Arkani-Hamed:2017jhn}
N.~Arkani-Hamed, T.-C. Huang and Y.-t. Huang, \enquote{{Scattering Amplitudes
  For All Masses and Spins}}, \href{http://arxiv.org/abs/1709.04891}{{\tt
  arXiv:1709.04891 [hep-th]}}.

\bibitem{Ochirov:2018uyq}
A.~Ochirov, \enquote{{Helicity amplitudes for QCD with massive quarks}},
  \href{http://dx.doi.org/10.1007/JHEP04(2018)089}{\emph{JHEP} \textbf{04}
  (2018) 089}, \href{http://arxiv.org/abs/1802.06730}{{\tt arXiv:1802.06730
  [hep-ph]}}.

\bibitem{Ballav:2020ese}
S.~Ballav and A.~Manna, \enquote{{Recursion relations for scattering amplitudes
  with massive particles}},
  \href{http://dx.doi.org/10.1007/JHEP03(2021)295}{\emph{JHEP} \textbf{03}
  (2021) 295}, \href{http://arxiv.org/abs/2010.14139}{{\tt arXiv:2010.14139
  [hep-th]}}.

\bibitem{Aoude:2019tzn}
R.~Aoude and C.~S. Machado, \enquote{{The Rise of SMEFT On-shell Amplitudes}},
  \href{http://dx.doi.org/10.1007/JHEP12(2019)058}{\emph{JHEP} \textbf{12}
  (2019) 058}, \href{http://arxiv.org/abs/1905.11433}{{\tt arXiv:1905.11433
  [hep-ph]}}.

\bibitem{Falkowski:2020aso}
A.~Falkowski and C.~S. Machado, \enquote{{Soft Matters, or the Recursions with
  Massive Spinors}}, \href{http://arxiv.org/abs/2005.08981}{{\tt
  arXiv:2005.08981 [hep-th]}}.

\bibitem{Boels:2011zz}
R.~H. Boels and C.~Schwinn, \enquote{{On-shell supersymmetry for massive
  multiplets}}, \href{http://dx.doi.org/10.1103/PhysRevD.84.065006}{\emph{Phys.
  Rev. D} \textbf{84} (2011) 065006},
  \href{http://arxiv.org/abs/1104.2280}{{\tt arXiv:1104.2280 [hep-th]}}.

\bibitem{johansson19}
H.~Johansson and A.~Ochirov, \enquote{{Double copy for massive quantum
  particles with spin}},
  \href{http://dx.doi.org/10.1007/JHEP09(2019)040}{\emph{JHEP.} \textbf{09}
  (2019) 040}.

\bibitem{DelDuca:1999rs}
V.~Del~Duca, L.~J. Dixon and F.~Maltoni, \enquote{{New color decompositions for
  gauge amplitudes at tree and loop level}},
  \href{http://dx.doi.org/10.1016/S0550-3213(99)00809-3}{\emph{Nucl. Phys. B}
  \textbf{571} (2000) 51}, \href{http://arxiv.org/abs/hep-ph/9910563}{{\tt
  arXiv:hep-ph/9910563}}.

\bibitem{Johansson:2015oia}
H.~Johansson and A.~Ochirov, \enquote{{Color-Kinematics Duality for QCD
  Amplitudes}}, \href{http://dx.doi.org/10.1007/JHEP01(2016)170}{\emph{JHEP}
  \textbf{01} (2016) 170}, \href{http://arxiv.org/abs/1507.00332}{{\tt
  arXiv:1507.00332 [hep-ph]}}.

\bibitem{Ochirov:2019mtf}
A.~Ochirov and B.~Page, \enquote{{Multi-Quark Colour Decompositions from
  Unitarity}}, \href{http://dx.doi.org/10.1007/JHEP10(2019)058}{\emph{JHEP}
  \textbf{10} (2019) 058}, \href{http://arxiv.org/abs/1908.02695}{{\tt
  arXiv:1908.02695 [hep-ph]}}.

\bibitem{Maltoni:2002mq}
F.~Maltoni, K.~Paul, T.~Stelzer and S.~Willenbrock, \enquote{{Color Flow
  Decomposition of QCD Amplitudes}},
  \href{http://dx.doi.org/10.1103/PhysRevD.67.014026}{\emph{Phys. Rev. D}
  \textbf{67} (2003) 014026}, \href{http://arxiv.org/abs/hep-ph/0209271}{{\tt
  arXiv:hep-ph/0209271}}.

\bibitem{Melia:2015ika}
T.~Melia, \enquote{{Proof of a new colour decomposition for QCD amplitudes}},
  \href{http://dx.doi.org/10.1007/JHEP12(2015)107}{\emph{JHEP} \textbf{12}
  (2015) 107}, \href{http://arxiv.org/abs/1509.03297}{{\tt arXiv:1509.03297
  [hep-ph]}}.

\bibitem{Lazopoulos:2021mna}
A.~Lazopoulos, A.~Ochirov and C.~Shi, \enquote{{All-multiplicity amplitudes
  with four massive quarks and identical-helicity gluons}},
  \href{http://arxiv.org/abs/2111.06847}{{\tt arXiv:2111.06847 [hep-th]}}.

\bibitem{Dixon:2010ik}
L.~J. Dixon, J.~M. Henn, J.~Plefka and T.~Schuster, \enquote{{All tree-level
  amplitudes in massless QCD}},
  \href{http://dx.doi.org/10.1007/JHEP01(2011)035}{\emph{JHEP} \textbf{01}
  (2011) 035}, \href{http://arxiv.org/abs/1010.3991}{{\tt arXiv:1010.3991
  [hep-ph]}}.

\bibitem{PhysRev.140.B516}
S.~Weinberg, \enquote{Infrared Photons and Gravitons},
  \href{http://dx.doi.org/10.1103/PhysRev.140.B516}{\emph{Phys. Rev.}
  \textbf{140} (1965) B516}.

\bibitem{Casali:2014xpa}
E.~Casali, \enquote{{Soft sub-leading divergences in Yang-Mills amplitudes}},
  \href{http://dx.doi.org/10.1007/JHEP08(2014)077}{\emph{JHEP} \textbf{08}
  (2014) 077}, \href{http://arxiv.org/abs/1404.5551}{{\tt arXiv:1404.5551
  [hep-th]}}.

\bibitem{Strominger:2013lka}
A.~Strominger, \enquote{{Asymptotic Symmetries of Yang-Mills Theory}},
  \href{http://dx.doi.org/10.1007/JHEP07(2014)151}{\emph{JHEP} \textbf{07}
  (2014) 151}, \href{http://arxiv.org/abs/1308.0589}{{\tt arXiv:1308.0589
  [hep-th]}}.

\bibitem{Hodges:2009hk}
A.~Hodges, \enquote{{Eliminating spurious poles from gauge-theoretic
  amplitudes}}, \href{http://dx.doi.org/10.1007/JHEP05(2013)135}{\emph{JHEP}
  \textbf{05} (2013) 135}, \href{http://arxiv.org/abs/0905.1473}{{\tt
  arXiv:0905.1473 [hep-th]}}.

\end{thebibliography}

\end{document}